\documentclass{emulateapj}
\usepackage{txfonts}
\usepackage{graphicx}
\usepackage{subfigure}
\usepackage{multirow}
\usepackage{url}
   
\newcommand{\lsim}{\mathrel{\hbox{\rlap{\lower.55ex\hbox{$\sim$}} \kern-.3em \raise.4ex \hbox{$<$}}}}
\newcommand{\gsim}{\mathrel{\hbox{\rlap{\lower.55ex\hbox{$\sim$}} \kern-.3em \raise.4ex \hbox{$>$}}}}
\newcommand{\beq}{\begin{equation}}
\newcommand{\eeq}{\end{equation}}
\newcommand{\beqa}{\begin{eqnarray}}
\newcommand{\eeqa}{\end{eqnarray}}
\newcommand{\drm}{\mathrm{d}}
\newcommand{\lcdm}{$\Lambda$CDM }

\newcommand{\ds}{d_\mathrm{S}}
\newcommand{\dls}{d_\mathrm{LS}}
\newcommand{\dl}{d_\mathrm{L}}
\newcommand{\vt}{v_\mathrm{T}}
\newcommand{\sigv}{\sigma_v}
\newcommand{\rhovir}{\bar{\rho}_\mathrm{vir}}
\newcommand{\mvir}{M_\mathrm{vir}}
\newcommand{\rvir}{R_\mathrm{vir}}
\newcommand{\Om}{\Omega_\mathrm{M}}
\newcommand{\Omo}{\Omega_\mathrm{M0}}
\newcommand{\mearth}{\,M_\oplus}
\newcommand{\msun}{\,M_\odot}
\newcommand{\mbd}{m_\mathrm{bd}}
\newcommand{\thetaEiso}{\theta_\mathrm{E}^\mathrm{SIS}}
\newcommand{\mtwo}{M_\mathrm{2D}}
\newcommand{\mr}{M_\mathrm{0.1pc}}
\newcommand{\Atot}{A_\mathrm{tot}}
\newcommand{\tobs}{t_\mathrm{obs}}
\newcommand{\nsub}{n_\mathrm{sub}}
\newcommand{\smin}{$\rm\it S_\mathrm{min} \rm$\,}
\newcommand{\uas}{$\mu$as\,}

\shorttitle{Local DM Subhalo Lensing}
\shortauthors{Erickcek \& Law}

\begin{document}

\title{Astrometric Microlensing by Local Dark Matter Subhalos}

\author{Adrienne L. Erickcek\altaffilmark{1,2} and Nicholas M. Law\altaffilmark{3}}
\altaffiltext{1}{Canadian Institute for Theoretical Astrophysics, University of Toronto, 60 St. George Street, Toronto, Ontario M5S 3H8, Canada; \mbox{erickcek@cita.utoronto.ca}}
\altaffiltext{2}{Perimeter Institute for Theoretical Physics, 31 Caroline St. N, Waterloo, Ontario N2L2Y5, Canada}
\altaffiltext{3}{Dunlap Fellow, Dunlap Institute for Astronomy and Astrophysics, University of Toronto, 50 St. George Street, Toronto ON M5S 3H4, Canada}

\begin{abstract}
High-resolution N-body simulations of dark matter halos indicate that the Milky Way contains numerous subhalos.  When a dark matter subhalo passes in front of a star, the light from that star will be deflected by gravitational lensing, leading to a small change in the star's apparent position.  This astrometric microlensing signal depends on the inner density profile of the subhalo and can be greater than a few microarcseconds for an intermediate-mass subhalo ($\mvir \gsim 10^4 \msun$) passing within arcseconds of a star.  Current and near-future instruments could detect this signal, and we evaluate the Space Interferometry Mission's (SIM's), Gaia's, and ground-based telescopes' potential as subhalo detectors.   We develop a general formalism to calculate a subhalo's astrometric lensing cross section over a wide range of masses and density profiles, and we calculate the lensing event rate by extrapolating the subhalo mass function predicted by simulations down to the subhalo masses potentially detectable with this technique. We find that, although the detectable event rates are predicted to be low on the basis of current simulations, lensing events may be observed if the central regions of dark matter subhalos are more dense than current models predict ($\gsim$$1 \msun$ within 0.1 pc of the subhalo center).  Furthermore, targeted astrometric observations can be used to confirm the presence of a nearby subhalo detected by gamma-ray emission.  We
show that, for sufficiently steep density profiles, ground-based adaptive optics astrometric techniques could be capable of detecting intermediate-mass subhalos at distances of hundreds of parsecs, while SIM could detect smaller and more distant subhalos.
\end{abstract}

\keywords{astrometry -- dark matter -- Galaxy:halo -- gravitational lensing:micro}

\maketitle
\section{Introduction}
Numerical simulations of dark matter halos have revealed the presence of numerous subhalos over a wide range of masses extending down to the simulations' resolution limits (e.g., \citealt{Ghigna98, Ghigna00, Klypin99, Moore99,  Kravtsov04, Gao04, Reed05, ViaLactea107, Aquarius08, ViaLactea208}). This substructure is the remnant of hierarchical structure formation; as halos merge to form larger structures, the inner portions of the ancestor halos become subhalos.  If all halos leave subhalo remnants, then the subhalo mass function may extend to masses far smaller than the resolution limit of any simulation \citep{Hu00, Chen01, Profumo06, DMS05}.

Although subhalos may be destroyed by gravitational interactions with the host halo, other subhalos, and stars, there are indications that their dense inner regions survive to the present day \citep{HNTS03, Kazantzidis04, VTG05, Read06, BDE06, Zhao07, GG07,  GGMDS07, SKM10, IME10}.  
Moreover, high-resolution simulations of halos similar to the Milky Way's host suggest that subhalos with masses greater than $10^6\msun$ are present at the Solar radius \citep{Aquarius08, ViaLactea208}.   Although simulations indicate that these large subhalos are significantly disrupted by the Galactic disk \citep{DSHK10, RSHH10}, the resolution is not sufficient to determine the fate of the subhalos' innermost regions.  It is therefore possible that numerous subhalos are located within a few kiloparsecs of the Solar System, with profound implications for both direct detection of the dark matter particle \citep{Kamion08} and indirect detection through its annihilation signature (e.g., \citealt{Bergstrom98, CM00, Stoehr03, DKM07, Ando2009, KKK10}).

Unfortunately, both indirect and direct detection of dark matter continues to be elusive, and gravity still provides the only uncontested evidence for dark matter.  Gravitational lensing is an especially powerful tool in the study of dark matter substructure; subhalos within our galaxy could be detected through their effects on signals from millisecond pulsars \citep{SHF07}, and substructure in lensing galaxies has several observational signatures.  Subhalos have been proposed as the origin of observed flux-ratio anomalies between the multiple images of strongly lensed quasars \citep{Mao98, MM01,Chiba02, DK02, KD04}; they can also alter the time delays between these images \citep{KM09, CKN10} and their separations \citep{Koopmans02, CRDT07, WFF08, More09}.  If the source is extended, then subhalos can distort the image's shape and surface brightness \citep{Metcalf02, IC05,IC05b, Koopmans05, VK09a, VK09b}.  Intriguingly, the observed lensing anomalies can only be explained if the central regions of the lensing galaxies contain significantly more substructure than is predicted by n-body simulations \citep{Mao04, Amara06, Maccio06, Xu09, Xu10}, although it has been suggested that these studies use atypical lensing galaxies \citep{Bryan08, Jackson10}.  

Individual subhalos within a lensing galaxy could be detected if they strongly lens one quasar image, splitting it into a closely spaced pair of images \citep{Yonehara03, IC05}.  Unfortunately, the diffuse nature of dark matter subhalos implies that they have small Einstein radii \citep{Zackrisson08}.  Consequently, the split images are resolvable only for the largest subhalos, making it unlikely to find a large enough subhalo with a sufficiently small impact parameter to detectably split a quasar image \citep{Riehm09}.  

In this paper, we consider a different way to find subhalos through gravitational lensing: instead of looking for split images, we investigate how the astrometric deflection of an image changes as a subhalo moves.  This astrometric microlensing approach has two advantages over strong lensing.  First, it is much easier to measure a change in the position of the centroid of an image than it is to resolve an image pair into distinct sources. The minimal separation required to resolve a pair of images is limited to approximately the resolution of the telescope ($\gsim25$ mas), whereas at high signal-to-noise ratio (S/N), the position of the centroid of the image can be measured with hundreds of times higher precision.  Second, strong lensing only occurs when the angular separation between the source and the lens is smaller than the lens's Einstein angle, while astrometric deflection is detectable for far larger impact parameters.   The disadvantage associated with astrometric microlensing is that it must be a dynamical event because we do not know the true position of the source.  As the lens moves relative to a background star, the star's image will move as well, and that is the detectable signature of astrometric microlensing.  We are therefore constrained to local subhalos with significant proper motions.     

Astrometric lensing signals from dark matter subhalos are necessarily small
because subhalos are diffuse, and astrometry has only very recently
progressed to the point that an astrometric dark matter search is
feasible. The rapid development of 1-100 $\mu$as astrometric precision
has been driven by a wide variety of fields -- for example, following
orbits in the galactic center (e.g., \citealt{Lu2009, Gillessen2009});
astrometric detection of planets (e.g., \citealt{Unwin2008, Law2009,
Malbet2009}); accurate parallax determination
(e.g., \citealt{Henry2009,Subasavage2009}); and the determination of stellar orbits (e.g.,
\citealt{Miniak2009, Pravdo2006, Konopacky2010, Ireland2008,
Dupuy2009}).

New space-based astrometric missions such as Gaia
\citep{Lindegren2008} and SIM \citep{Unwin2008} are opening the
possibility of ultra-high-precision all-sky and targeted searches.
Targeted ground-based astrometry is already capable of 100$\mu$as
precision in arcminute-sized fields, and new larger telescopes will
significantly improve that precision \citep{Cameron2009}, while even
higher precisions are possible on bright stars (e.g.,
\citealt{Muterspaugh2006, vanBelle2008}). 

These technical advances have inspired considerable interest in astrometric microlensing by stars and dark compact objects \citep{Walker95, Paczynski95, Paczynski98, Miralda96,Gould00,Gaudi2005}, and by baryonic clouds \citep{Takahashi03,Lee10}.   In this paper,
we investigate if these instruments can also be used to search for dark matter
subhalos.  We find that, for standard dark matter models, observing a subhalo lensing event rate during a blind astrometric survey is highly unlikely.  If the central regions of dark matter subhalos are denser than expected, however, the lensing event rate can be significantly enhanced.   We also explore the possibility of using astrometric lensing to confirm the presence of a subhalo detected through its gamma-ray emission.  We find that ground-based telescopes could detect lensing by a nearby ($\sim$50 pc) subhalo with a post-stripping mass greater than $1000\msun$, while SIM could probe these subhalos at greater distances ($\sim$100 pc) and detect nearby subhalos with one-tenth of this mass.  

This article is organized as follows. In Section \ref{sec:astrometric_sigs}, we describe the image motion induced by subhalo lenses with a variety of different density profiles, namely, a singular isothermal profile, the NFW profile, and a generalized power-law density profile. In Section \ref{Sec:strategy}, we develop an astrometric observing strategy designed to reliably detect subhalo lensing while rejecting possible false-positive detections.  In Section \ref{sec:x-sections}, we calculate the areas of sky over which particular subhalo lenses are detectable and determine the all-sky subhalo lensing event rates for several models of dark matter substructure.  In Section \ref{sec:detection_prospects}, we evaluate current and planned astrometric survey capabilities in the context of detecting lensing from a subhalo in both all-sky and targeted searches. We summarize our findings and conclude in Section \ref{sec:summary}.    To evaluate the lensing signatures and event rates, we developed models of the concentration-mass relation and the mass function for local subhalos based on the findings of the Aquarius simulations \citep{Aquarius08}; these models are presented in the appendix.

\section{Astrometric Signatures of Lensing by Subhalos}
\label{sec:astrometric_sigs}
The shapes of the intermediate-mass subhalos that we will be considering cannot be probed by current numerical simulations; to simplify our analysis we will assume that the subhalos are spherically symmetric.  Numerical simulations can only probe the shapes of the largest subhalos with masses greater than $10^8 \msun$ \citep{Kuhlen07, Knebe10}.  While these subhalos are triaxial, they tend to be more spherical than their host halos.  In a simulation that includes baryonic physics, \citet{Knebe10} found that subhalos located within half of the virial radius of the host halo have nearly spherical mass distributions in their innermost resolved regions, with a median minor-to-major axis ratio of 0.83.  We will see that astrometric microlensing is only sensitive to the density profile near the center of the subhalo ($r<0.1$ pc); as long as the inner region of the subhalo is nearly spherically symmetric, the subhalo's triaxiality is irrelevant for our analysis, and we do not expect deviations from spherical symmetry to significantly affect our results.  

We will also assume that the dark matter subhalo is transparent and that it contains no stars.  The presence of a star in the center of the subhalo would change its lensing signature if the star's mass is comparable to the dark matter mass in the central 0.1 pc of the subhalo, but stars further from the center would not have an effect.  Furthermore, the discrepancy between the number of intermediate-mass subhalos seen in simulations and the number of dwarf spheroidals observed in the Milky Way, known as the ``missing satellite problem" \citep{Klypin99b, Moore99}, indicates that stars are rare in subhalos with masses less than $10^6\msun$ \citep{Madau08}.  
Finally, we will assume that the subhalo's diameter is small compared to both the distance $\dl$ between the lens and the observer and the distance $\dls$ between the lens and the star.   

\begin{figure*}[tb]
 \centering
 \resizebox{\textwidth}{!}
 {
      \includegraphics{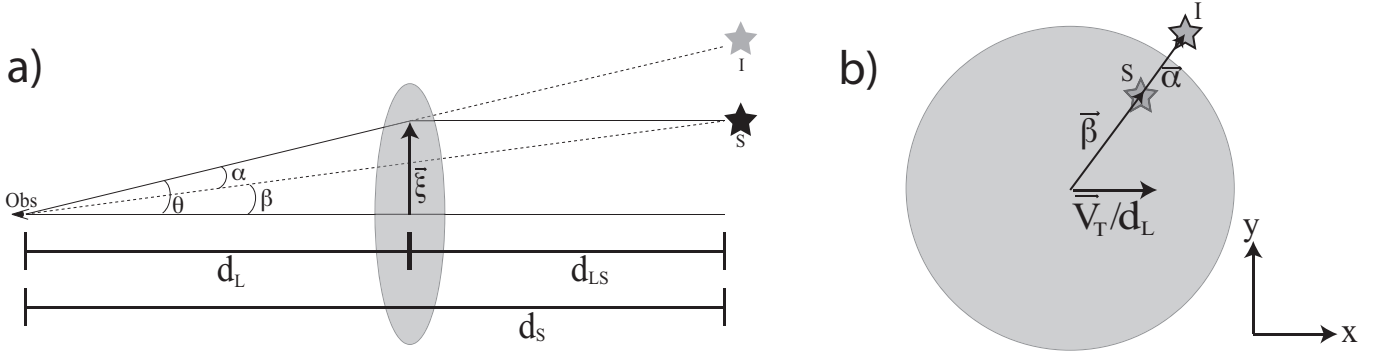}
 }
\caption{a) Diagram showing the position of the source star (in black), its image (in gray), and the lens (gray ellipse). We will generally assume that $\alpha\ll\beta$ so that the ray's impact parameter in the lens plane ($\xi\equiv d_L\theta$) may be approximated as $\xi \simeq d_L \beta$.  b) The same lensing system viewed as projected on the sky.  The center of the lens is moving with velocity $\vt$ along the $x$-axis.}
\label{Fig:lensdiagram}
\end{figure*}

Figure \ref{Fig:lensdiagram} shows a schematic view of lensing by an extended transparent object like a dark matter subhalo. 
When a light ray passes through a spherically symmetric thin lens, the image of the star is shifted from its true position by an angle 
\beq
\vec{\alpha} = \frac{\dls}{\ds} \left[\frac{4GM_\mathrm{2D}(\xi)}{\xi}\right]\hat{\xi},
\label{alpha}
\eeq
where $\ds$ is the distance between the observer and the star, $\xi$ is the distance between the center of the lens and the star's image in the lens plane ($\vec{\xi} \equiv \dl \vec{\theta}$), and $\hat{\xi} \equiv \vec{\xi}/\xi$ points from the lens to the star.  Throughout this work, we set the speed of light $c=1$.  The mass $M_\mathrm{2D}(\xi)$ in Eq.~(\ref{alpha}) is the mass enclosed by the cylinder interior to $\xi$ and is obtained by integrating the projected surface mass density $\Sigma$ over the area of the circle with radius $\xi$.   

As the subhalo moves relative to the background star, the angle $\vec\beta$ that extends from the lens to the star will change, and the position of the star's image will change accordingly. We take the star to be fixed at the origin of an $xy$ coordinate system on the sky, and we define the $x$-axis to be parallel to the subhalo's transverse velocity $\vt$, as shown in the right panel of Fig.~\ref{Fig:lensdiagram}.  The vertical component of $\vec\beta$ is therefore fixed [$\beta_y(t) = \beta_{y,0}$], and 
\beq
\beta_x(t)= \beta_{x,0} - 4.2^{\prime\prime}\,\left(\frac{\vt}{200\,\mathrm{km/s}}\right)\left(\frac{\mathrm{50\,pc}}{\dl}\right)\left(\frac{t}{5\,\mathrm{yr}}\right),
\label{beta}
\eeq
where $\beta_{x,0}$ is the value of $\beta_x$ at $t=0$.  We see that a nearby subhalo will move several arcseconds during a multi-year observational period.  

For the subhalos we consider, the deflection angle $\alpha$ will be on the order of microarcseconds.  Since $\beta$ changes by several arcseconds over the course of the observation, $\beta\gg\alpha$ for most of the observational period.  We will further assume that $\beta_{y,0}\gg \alpha$ so that we are always considering the weak-lensing regime with $\beta\gg\alpha$.  We verify in Appendix \ref{Sec:multiandtrunc} that this condition is satisfied for all subhalo lensing scenarios, provided that $\dl \ll 1000$ kpc.  This confirms that we are firmly in the weak lensing regime as long as we only consider subhalos in our local group.  In this case, there is only one image of the star, and it is always located on the line connecting the lens position to the star's position, with the star between the image  and the lens.  We will use the $\beta\gg\alpha$ assumption to simplify the lens equation by approximating $\vec\beta \simeq \vec\theta$.  In this case, $\vec \xi$ may be approximated as $\vec\xi \simeq \dl \vec\beta$, and Eq.~(\ref{alpha}) becomes a simple equation for the deflection angle $\vec\alpha$ in terms of the impact parameter $\vec\beta$.  In the following subsections, we will use Eq.~(\ref{alpha}) to show how the path taken by the star's image during a subhalo transit depends on the subhalo's density profile.  

\subsection{Singular isothermal sphere}
The density profile for a singular isothermal sphere (SIS) is
\beq
\rho(r) = \frac{\sigv^2}{2\pi G r^2},
\eeq
where $\sigv$ is the velocity dispersion of the halo.  Although numerical simulations indicate that large dark matter subhalos without baryons do not have this steep a density profile \citep{Aquarius08, ViaLactea208}, we consider the SIS case in detail because it simply illustrates key features that are shared by the astrometric lensing signatures from dark matter halos with shallower profiles.  

The two-dimensional enclosed mass for an infinite SIS is $\mtwo(\xi) = \pi\sigv^2\xi/G$.  Since $\mtwo$ depends linearly on $\xi$, Eq.~(\ref{alpha}) reveals that $\alpha$ is independent of the separation between the lens and the star.   The deflection angle is always the Einstein angle of the SIS:
\beqa
\thetaEiso &=& \left(1-\frac{\dl}{\ds}\right)\,4\pi\sigv^2, \label{thetaEiso}\\
&=& 10\, \mu\mathrm{as}\, \left( \frac{\sigv}{0.6\, \mathrm{km/s}}\right)^2\left(1-\frac{\dl}{\ds}\right).
\eeqa
There are two images, with $\vec{\alpha} = \pm \thetaEiso \hat{\beta}$, only if $\beta < \thetaEiso$.  We will only consider larger impact parameters, in which case there is only one image, with $\vec \alpha = \thetaEiso \hat{\beta}$.   As the SIS moves relative to the star, the direction of the deflection angle changes.  For an infinite SIS moving from the distant left to the distant right, the image starts $\thetaEiso$ to the right of the star's true position and then traces a half-circle with radius $\thetaEiso$ until it ends $\thetaEiso$ to the left of the star's true position.

For an SIS, the mass enclosed in a sphere of radius $R$ is proportional to $R$; if the SIS has infinite extent, then its mass is infinite.  It is customary to characterize an SIS by its virial mass: the mass contained in a sphere with mean density equal to the virial density $\rhovir$.  \citet{BN98} provide a fitting formula for the virial density in a flat \lcdm universe,
\beqa
\rhovir &\equiv& \left(18\pi^2+82[\Om(z)-1]-39[\Om(z)-1]^2\right)\rho_\mathrm{crit}, \\
\Om(z) &=& \frac{\Omo(1+z)^3}{\Omo(1+z)^3+1-\Omo}, \\
\rho_\mathrm{crit}(z)&=& \left(0.0924\, \frac{\mearth}{\mathrm{pc}^3} \right) h^2 \left[\Omo(1+z)^3+1-\Omo\right],
\eeqa
where $H_0 \equiv 100 h \, \mathrm{km \,s^{-1} \,Mpc^{-1}}$ and $\Omo$ is the present-day matter density divided by the critical density.  We will use standard cosmological parameters: $h=0.7$ and $\Omo=0.3$.
The virial density is 4.6 $\mearth$ pc$^{-3}$ at redshift zero and it increases monotonically with redshift.  
The velocity dispersion $\sigv$ in terms of the virial mass is
\beq
\sigv^2 = G \left(\frac{\pi\rhovir(z_v)\mvir^2}{6}\right)^{1/3},
\eeq
where $z_v$ is the redshift at which the halo's virial mass is evaluated.  To facilitate comparisons with N-body simulations, we take $z_v=0$ in our calculations, but we note that increasing $z_v$ would make the subhalos denser and would enhance their lensing signals.  Inserting this expression into Eq.~(\ref{thetaEiso}) gives
\beq
\thetaEiso = \left(1.1 \, \mu\mathrm{as}\right) \left(1-\frac{d_L}{d_S}\right)\left(\frac{\mvir}{10^4 \msun}\right)^{2/3}\left(\frac{\rhovir(z_v)}{4.6 \mearth \, \mathrm{pc}^{-3}}\right)^{1/3}.
\label{thetaEiso2}
\eeq

These properties describe an SIS in isolation.  Once a subhalo is accreted by a larger halo, the outer tails of its density profile are stripped of mass.  Numerical simulations indicate that a subhalo in our Galaxy may lose between 99\% and 99.9\% of its initial mass due to tidal stripping from the smooth component of the halo \citep{HNTS03, Kazantzidis04, VTG05, Read06}, and stars will further strip the outer portions of subhalos \citep{BDE06,Zhao07, GG07,  GGMDS07, SKM10, IME10}.   We will deal with this truncation by defining a truncation radius $R_t$ and setting $\rho=0$ for $R>R_t$.  The mass contained within $R_t$ is the mass of the surviving subhalo $M_t$.  We will describe the tidal stripping with the parameter $\mbd \equiv M_t/\mvir$, where $\mvir$ is the original virial mass of the subhalo, evaluated at redshift $z_v$.  The truncation radius is then given by
\beq
R_t = \left(0.56 \, \mathrm{pc}\right)\left(\frac{\mbd}{0.001}\right)\left(\frac{\mvir}{10^4 \msun}\right)^{1/3}\left(\frac{4.6 \mearth \, \mathrm{pc}^{-3}}{\rhovir(z_v)}\right)^{1/3}.
\eeq
The angular size of the truncated halo is
\beq
\theta_t = 0.64^\circ \left(\frac{\mbd}{0.001}\right)\left(\frac{50\,\mathrm{pc}}{d_L}\right)\left(\frac{\mvir}{10^4 \msun}\right)^{1/3}\left(\frac{4.6 \mearth \, \mathrm{pc}^{-3}}{\rhovir(z_v)}\right)^{1/3}.
\eeq
Thus we see that $\thetaEiso \ll \theta_t$ for all subhalos of interest.

For a truncated SIS, the two-dimensional enclosed mass is
\beq
\mtwo (\xi < R_t) = \frac{2\sigv^2}{G}\left[\xi \tan^{-1}\sqrt{\frac{R_t^2}{\xi^2}-1}+R_t-\sqrt{R_t^2-\xi^2}\right]
\label{mtwoiso}
\eeq
and $\mtwo = M_t$ for $\xi \geq R_t$.  If $\xi \geq R_t$, then the deflection angle is the same as for a point mass with Einstein angle
\beq
\theta_E^\mathrm{PM} \equiv \sqrt{\frac{4GM_t \dls}{\dl \ds}} = \sqrt{\frac{2}{\pi}\thetaEiso \theta_t}.
\eeq
Thus we see that $\thetaEiso \ll \theta_t$ implies that $\theta_E^\mathrm{PM} \ll \theta_t$.  Therefore we may approximate the position of the brightest image as $\alpha =  (\theta_E^\mathrm{PM})^2/\beta$ for $\beta > \theta_t$.  When we insert $\mtwo$ from Eq.~(\ref{mtwoiso}) into Eq.~(\ref{alpha}) for $\alpha$ and assume that $\alpha \ll \beta$, we find that the deflection angle for lensing by a truncated SIS is
\beqa
\vec\alpha &=&  \frac{2}{\pi}\thetaEiso\,\hat\beta \left\{\begin{array}{ll} {\cal F}\left(\frac{\beta}{\theta_t}\right) 
& \mathrm{for} \quad \thetaEiso<\beta<\theta_t \\
\frac{\theta_t}{\beta} & \mathrm{for}\quad \beta>\theta_t 
\end{array}\right.,\\
{\cal F}(x) &\equiv& \tan^{-1}\sqrt{\frac{1}{x^2}-1}+\frac{1}{x}-\sqrt{\frac{1}{x^2}-1} \nonumber.
\eeqa

\begin{figure}
\centering
\includegraphics[width=8.2cm]{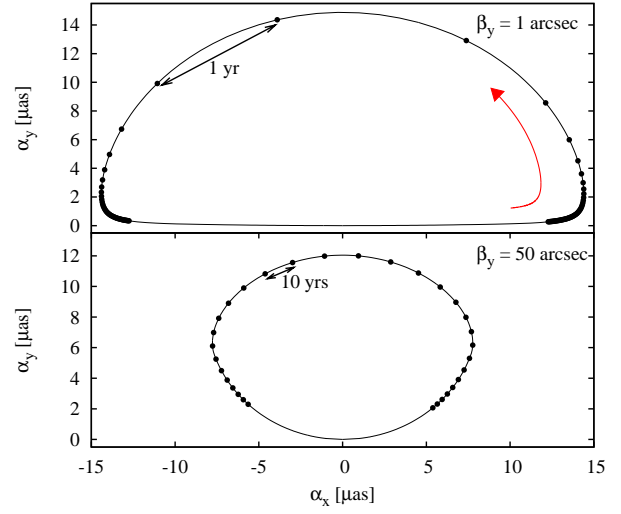}
\caption{The deflection angle for a star at 5 kpc as an SIS subhalo 50 pc away moves at a velocity of 200 km/s from the far left to the far right.  This subhalo's velocity dispersion is $0.72$ km/s, which corresponds to a virial mass of $5\times10^5 \, \msun$, but it has been stripped to a radius of 0.02 pc and contains only 5 $\msun$.  In the top plot, the subhalo center passes 1 arcsecond below the star, and the circles show the image's position every year for the 100 years surrounding the time of closest approach.  In the bottom plot, the subhalo center passes 50 arcseconds below the star, and the circles show the image's position every 10 years for the 300 years surrounding the time of closest approach.}
\label{Fig:SISimagepath}
\end{figure}

Figure \ref{Fig:SISimagepath} shows how the image of a fixed star moves as a truncated SIS subhalo passes below the star's true position on the sky.  The lens's Einstein angle is $\thetaEiso=15\mu$as, which corresponds to $\sigv=0.72$ km/s and $\mvir=5\times10^5 \msun$.  The image motion is highly sensitive to the ratio $\vt/\dl$ because this ratio determines how $\vec{\beta}$ changes with time [see Eq.~(\ref{beta})].  If $\vt/\dl$ is decreased, then the change in $\vec\beta$ during a set time interval is decreased, and the image motion slows down.  Throughout this work, we adopt $\vt = 200$ km/s because that is the typical velocity of a dark matter particle in the halo (e.g., \citealt{Drukier86, Xue08}).  With this velocity, we will see that a detectable subhalo must have $\dl \lsim 100$ pc, and we adopt $\dl = 50$ pc as our fiducial lens distance.  Provided that $\dl \ll \ds$, the distance to the source has a minimal impact on the image motion because $\ds$ only enters through the factor $(1-\dl/\ds)$ in $\thetaEiso$ [see Eq.~(\ref{thetaEiso2})].  We use $\ds=5$ kpc as our fiducial value because this is a reasonable distance to a target star; changing $\ds$ to any value greater than 1 kpc would have no noticeable effect on the image motion.  To illustrate the effects of subhalo truncation in Fig.~\ref{Fig:SISimagepath}, we assume that the subhalo has been extremely stripped by close encounters with stars so that its radius is 0.02 pc ($\theta_t = 85^{\prime\prime}$), which implies that $M_t = 5\msun$.   We see that the image motion consists of three stages: as the edge of the subhalo approaches from the far left, the image very slowly moves rightward until the star is behind the subhalo, then the image rapidly traces an arc as the subhalo center passes by the star, and finally the star slowly returns to its true position as the subhalo moves off to the right.

\begin{figure}
\centering
\includegraphics[width=8cm]{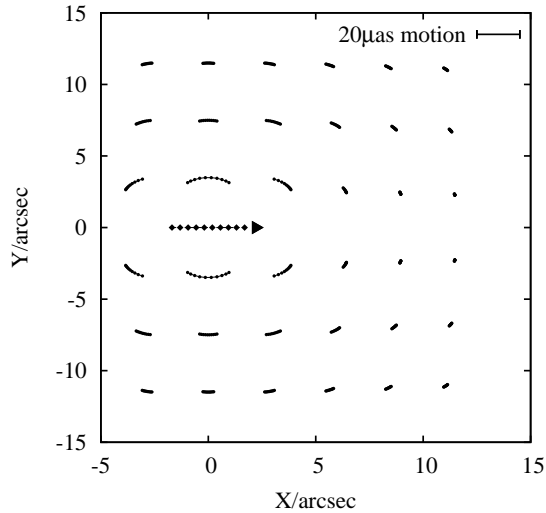}
\caption{The deflection induced over 4 years by a moving SIS lens with the same properties as in Fig.~\ref{Fig:SISimagepath}.  The path of the lens is depicted by a dotted arrow. To show the image trajectories, the image motion is exaggerated a factor of $10^6$ relative to the star's positions; a scale bar corresponding to 20$\mu$as motion is shown. Twenty equally spaced measurement points over the 4-year observational period are shown for each curve. Note that the stars closest to the lens position undergo much more rapid position changes.}
\label{Fig:SISimagegrid}
\end{figure}

The impact parameter $\beta_y$ determines how quickly the image moves as the subhalo passes by the star.  In the top half of Fig. \ref{Fig:SISimagepath} the lens impact parameter is $1^{\prime\prime}$, and the image rapidly traces out a semi-circle of radius $\thetaEiso$ during the few years surrounding the time of closest approach, just as if the lens had infinite extent.  The effect of the SIS's truncation is more apparent in the bottom half of  Fig. \ref{Fig:SISimagepath}, where $\beta_y=50^{\prime\prime}$; the image's trajectory is closer to a circle, and it will become more and more circular as $\beta_y$ increases.  The image transverses this circle very slowly, taking 10 years to move 2 $\mu$as, in contrast to the image in the top panel, which moves nearly 30 $\mu$as in only 5 years.   Thus we see that the only detectable portion of the image's path in the sky is the period surrounding the moment of closest approach between the star and the lens, and a small impact parameter is required to make the image move significantly over the course of a few years.  Figure \ref{Fig:SISimagegrid} further illustrates the necessity of a small impact parameter by showing how the images of stars at different positions relative to the lens move over the course of 4 years; only the stars along the lens's path with $\beta_y \lsim 2^{\prime\prime}$ have images that are significantly moved during the observational period.   For stars that are this close to the center of the subhalo, with $\beta \ll \theta_t$, the truncation of the density profile does not affect the image trajectories, as seen in the top panel of Fig. \ref{Fig:SISimagepath}.  We will therefore assume that $\beta \ll \theta_t$ for all interesting lensing scenarios and ignore the subhalo's truncation when considering other density profiles. 

\subsection{NFW density profile}
The NFW profile,
\beq
\rho(r) = \frac{\rho_s}{\left(\frac{r}{r_s}\right)\left(1+\frac{r}{r_s}\right)^2},
\eeq
was found to be a universally good fit to the density profiles of galaxy and cluster halos in early numerical simulations \citep{NFW96, NFW97}.  
The two-dimensional enclosed mass for a subhalo with virial mass $\mvir$ and an NFW density profile with concentration $c \equiv \rvir/r_s$ is 
\beqa
\mtwo(\xi) &=&  \frac{\mvir}{\ln(1+c) - \frac{c}{1+c}} \,{\cal G}\left(\frac{\xi}{r_s}\right) \\
{\cal G}(x) &=& \ln\frac{x}{2}+\left\{\begin{array}{ll}
 \frac{1}{\sqrt{1-x^2}}\cosh^{-1}\frac{1}{x} & \mathrm{for} \quad x<1 \\
1 & \mathrm{for}\quad x=1\\
 \frac{1}{\sqrt{x^2-1}}\cos^{-1}\frac{1}{x} & \mathrm{for} \quad x>1
\end{array}\right.\nonumber
\eeqa
\citep{Bartelmann96, GK02}.  

From Eq.~(\ref{alpha}), we see that the magnitude of the deflection angle $\alpha$ is proportional to $\mtwo/\xi$.  For the NFW profile, $\alpha \propto \xi$ if $r\ll r_s$, and $\alpha \propto \xi^{-1}$ if $r\gg r_s$.  Therefore, as an NFW subhalo approaches a star, the deflection angle will increase until the star crosses the scale radius ($\xi \simeq r_s$), and then it will decrease until the star crosses the subhalo center, at which point it will begin to increase again until the star crosses $r_s$ on the other side of the subhalo.  In this sense, the scale radius of an NFW profile plays the same role as the truncation radius for a truncated SIS.  If the impact parameter is close to the scale radius ($\beta_y \simeq r_s/\dl$), then the image trajectory is roughly circular, and it resembles the bottom half of Fig.~\ref{Fig:SISimagepath}.  Unfortunately, the subhalos that are massive enough to deflect the star's image by several microarcseconds ($\mvir \gsim 10^4 \msun$) have large scale radii ($r_s \gsim 2$ pc for $c\lsim 100$); $\beta_y \simeq r_s/\dl$ is a large impact parameter, and the image position changes very slowly as the subhalo moves.  Moreover, just as with a truncated SIS lens, the reversal in the image's motion as the star crosses the scale radius ($\beta \simeq r_s/\dl$) is very slow, regardless of the impact parameter $\beta_y$.

\begin{figure}
\centering
\includegraphics[width=8.5cm]{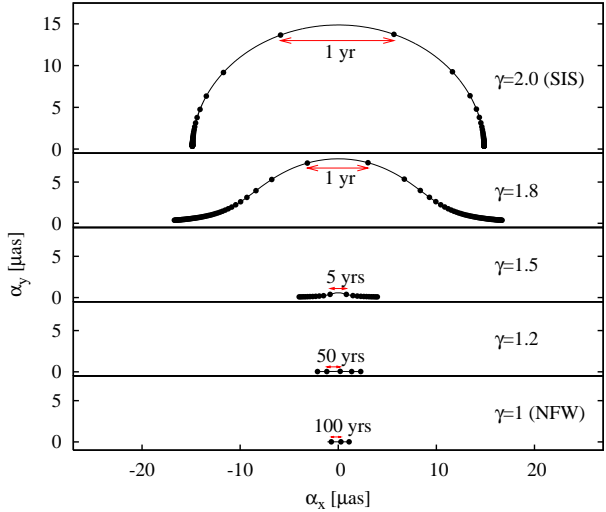}
\caption{The deflection angle generated by a moving lens with $\ds =5$ kpc, $\dl = 50$ pc and $\vt = 200$ km/s.  The virial mass of the lens is $5\times10^5\, \msun$, and its concentration is $ \rvir/ r_{-2}= 99$.  The inner density profile of the lens is given by $\rho \propto r^{-\gamma}$, and the different panels correspond to different values of $\gamma$.  The impact parameter is 1 arcsecond, and only the portion of the image path corresponding to the time surrounding the moment of closest approach between the image and lens is shown. Note that the image path becomes more linear and the image motion slows down considerably as $\gamma$ is decreased.}
\label{Fig:GENimagepath}
\end{figure}

As in the SIS case, the most promising lensing scenario occurs when the center of the NFW subhalo passes very close to the source.   The key difference is that $\alpha$ is nearly constant for $\xi\ll R_t$ if the lens is an SIS, which leads to the semi-circle image trajectory displayed in the top portion of Fig.~\ref{Fig:SISimagepath}.  For an NFW lens with $\xi \ll r_s$, the deflection angle is very small, as shown in the bottom panel of Fig.~\ref{Fig:GENimagepath}.   The NFW density profile leads to a no-win situation: if you decrease the impact parameter $\beta_y$ in order to enhance the change in the image's position over a set time period, the magnitude of the deflection decreases.  We are forced to conclude that astrometric lensing by subhalos is only detectable if the inner density profile is steeper than $\rho \propto r^{-1}$.

\subsection{Generalized density profile}
\label{Sec:genprofile}

We have seen that astrometric gravitational lensing by subhalos is only detectable if the center of the subhalo passes close to the star's position during the observational period, during which the subhalo moves about 0.001 pc (for a 5-year observational period).  Therefore, only the innermost portion of the subhalo is responsible for the astrometric lensing signature.  Unfortunately, very little is known about the intermediate-mass ($10 \msun \lsim M_t \lsim 10^6\msun$) subhalos that are capable of producing detectable astrometric lensing events.  High-resolution N-body simulations can probe the density profiles of only the largest ($M_t\gsim10^8\msun$) subhalos, and even these profiles are unresolved at $r\lsim 350$ pc \citep{Aquarius08, ViaLactea208}.  For these large subhalos, \citet{ViaLactea208} find that $\rho\propto r^{-1.2}$ in the innermost resolved regions, while \citet{Aquarius08} see $\rho\propto r^{-(1.2-1.7)}$ at their resolution limit for nine large subhalos, with no indication that the slope had reached a fixed central value.  Meanwhile, at the opposite end of the mass spectrum, \cite{DMS05} find that the first Earth-mass dark matter microhalos have steeper density profiles with $\rho\propto r^{-(1.5-2.0)}$ at redshift $z=26$, and higher-resolution simulations indicate that this steep profile extends to within 20 AU of the microhalo center \citep{IME10}.  

In light of this uncertainty, we consider a generic density profile
\beq
\rho(r) = \rho_0 \left(\frac{r}{r_0}\right)^{-\gamma}
\label{innerprofile}
\eeq
with $1<\gamma\leq2$.  We assume that a constant-density core, if present, is significantly smaller than our typical impact parameters of 0.001 pc, and we assume that the subhalo does not contain a black hole.  Larger cores would decrease the lensing signal while the presence of a black hole would enhance it by  adding a point mass and steepening the density profile \citep{BZS05, RG09}.  If we take this density profile as infinite when calculating the projected surface density $\Sigma$, we find that 
\beqa
\Sigma(\xi) &=& \sqrt{\pi} \,\rho_0 r_0\, \frac{\Gamma\left[0.5(\gamma-1)\right]}{\Gamma\left[0.5\gamma\right]} \left(\frac{\xi}{r_0}\right)^{1-\gamma}, \label{GenSigma}\\
\mtwo(\xi) &=& 2 \pi^{3/2}\,\left(\rho_0 r_0^3\right) \,\frac{\Gamma\left[0.5(\gamma-1)\right]}{(3-\gamma)\Gamma\left[0.5\gamma\right]} \left(\frac{\xi}{r_0}\right)^{3-\gamma} \label{GenMtwo},
\eeqa
where $\Gamma[x]$ is the Euler gamma function.  

Of course, this density profile does not extend to infinity; the subhalo's density profile will be truncated by tidal stripping, and it may also transition to a steeper power law, as in the case of an NFW profile.  If the density profile is truncated at $r=R_t$, then the surface density diverges from Eq.~(\ref{GenSigma}) as $\xi$ approaches $R_t$, but for $\xi\ll R_t$, Eqs.~(\ref{GenSigma}) and (\ref{GenMtwo}) are still good approximations.  For instance, if $\gamma = 1.5 \,(1.2)$, $\mtwo (\xi)$ for a subhalo truncated at $R_t$ is greater than 80\%  (50\%) the value given by Eq.~(\ref{GenMtwo}) if $\xi \leq 0.1 R_t$.  We will show in Appendix \ref{Sec:multiandtrunc} that detectible astrometric signatures are only produced if $\xi < 0.03$ pc, and Eq.~(\ref{GenMtwo}) is accurate to within $20\%$ for subhalos with $\gamma \geq 1.5$, $\mvir < 10^8 \msun$, and $R_t \gsim 0.1$ pc.  Furthermore, the lower bound on $R_t$ is significantly smaller for subhalos with  $\mvir \ll 10^8 \msun$.  We will therefore use Eq.~(\ref{GenMtwo}) and take $R_t \gsim 0.1$ pc as a conservative lower bound, although we note that the resulting deflections may be slightly overestimated, especially if $\gamma \lsim 1.2$.  As shown in Fig.~\ref{Fig:GENimagepath}, however, detecting a subhalo with $\gamma \lsim 1.2$ is challenging, and we conclude that Eq.~(\ref{GenMtwo}) is accurate to within $\sim20\%$ for subhalos of interest.

If a dark matter subhalo with a density profile given by Eq.~(\ref{innerprofile}) passes in front of a star, Eq.~(\ref{alpha}) tells us that 
\beq
\vec\alpha = \theta_\alpha \left(\frac{\xi}{r_0}\right)^{2-\gamma} \hat{\xi},\label{genalpha}
\eeq
where we have defined
\beq
\theta_\alpha \equiv 0.88\,\mu\mathrm{as}\,\left( \frac{\Gamma\left[0.5(\gamma-1)\right]}{2(3-\gamma)\Gamma\left[0.5\gamma\right]}\right)\left(1-\frac{d_L}{d_S}\right) \left(\frac{\mathrm{pc}}{r_0}\right)\left(\frac{\rho_0 r_0^3}{\msun}\right).
\label{the_alph}
\eeq
Like $\thetaEiso$, $\theta_\alpha$ depends on the distances to the lens and the source only through the factor $(1-\dl/\ds)$.   We also note that $\theta_\alpha$ is related to the Einstein angle $\theta_\mathrm{E}$:
\beq
\theta_\alpha = \theta_\mathrm{E}^{\gamma-1}\left(\frac{r_0}{\dl}\right)^{2-\gamma}.
\label{genThetaE}
\eeq  
We will continue to assume that $\alpha \ll \beta$ so that $\xi$ (see Fig.~\ref{Fig:lensdiagram}) is approximately equal to $\dl \beta$.

Equation (\ref{the_alph}) gives the magnitude of the deflection angle in terms of the parameters of the density profile $r_0$ and $\rho_0$, but this is not the most useful description of the subhalo.  Instead we characterize the subhalo by either its mass after tidal stripping ($M_t \equiv \mbd\mvir$) or the mass contained within a radius of 0.1 pc from the subhalo center ($\mr$).  Although $M_t$ is a more standard and intuitive description of the subhalo mass, using $\mr$ offers two advantages.  First, $\mr$ completely determines the deflection angle; without loss of generality, we can set $r_0=0.1$ pc, in which case
\beq
\theta_\alpha =8.8\,\mu\mathrm{as}\,\left( \frac{\Gamma\left[0.5(\gamma-1)\right]}{2(3-\gamma)\Gamma\left[0.5\gamma\right]}\right)\left(1-\frac{d_L}{d_S}\right)\left(\frac{3-\gamma}{4\pi}\right)\left(\frac{\mr}{\msun}\right).
\eeq
Second, $\mr$ is the portion of the subhalo's mass that is actually probed by astrometric microlensing because truncating the subhalo's density profile at $R_t=0.1$ pc does not affect its astrometric lensing signature.  Therefore, using $\mr$ to characterize the subhalo's mass allows us to consider subhalos that are more compact than standard virialized subhalos and makes it easy to apply our results to more exotic forms of dark matter substructure.  

To relate $\theta_\alpha$ to the virial mass of the subhalo, we have to specify the full density profile.  If $\gamma=2.0$, we will assume that the subhalo is a truncated SIS so that Eq.~(\ref{innerprofile}) holds out to the truncation radius of the subhalo.  In this case, Eq.~(\ref{genThetaE}) tells us that $\theta_\alpha = \thetaEiso$, and we can use Eq.~(\ref{thetaEiso2}) to evaluate $\theta_\alpha$.  If $\gamma \neq 2$, we will assume that the subhalo's full density profile prior to any tidal stripping was
\beq
\rho(r) = \frac{\rho_0}{\left(\frac{r}{r_0}\right)^\gamma\left(1+\frac{r}{r_0}\right)^{3-\gamma}},
\label{GenNFW}
\eeq
which reduces to Eq.~(\ref{innerprofile}) for $r \ll r_0$.  In this case, the virial mass does not uniquely determine $\theta_{\alpha}$, and we also have to specify the subhalo's concentration.  We define the concentration as $c\equiv \rvir/ r_{-2}$, where $r_{-2}$ is the radius at which $d\ln \rho/d\ln r = -2$.  For the profile given by Eq.~(\ref{GenNFW}), $r_{-2} = (2-\gamma)r_0$.  It follows that 
\beq
r_0 = 27\, \mathrm{pc} \,\left(\frac{1}{2-\gamma}\right)\left(\frac{94}{c}\right)\left(\frac{\mvir}{10^6 \msun}\right)^{1/3}\left(\frac{\rhovir(z_v)}{4.6 \mearth \, \mathrm{pc}^{-3}}\right)^{-1/3}.
\label{r0}
\eeq
We see that $r_0$ is larger than 5 pc for the subhalos we consider ($\mvir \geq 10^4 \msun$ and $\gamma \geq 1.2$), and we expect that local subhalos will be stripped to much smaller radii by encounters with stars.  

We can now derive how $\theta_\alpha$ depends on the subhalo's virial mass and concentration.  Recall from Eq.~(\ref{the_alph}) that $\theta_\alpha \propto \rho_0 r_0^2$.  For the profile given by Eq. (\ref{GenNFW}), this factor is related to the subhalo's concentration and virial mass through
\beqa
\left(\frac{\mathrm{pc}}{r_0}\right)\left(\frac{\rho_0 r_0^3}{\msun}\right) &=& 810\, \left(\frac{c}{94}\right) \left(\frac{\mvir}{10^6 \msun}\right)^{2/3}\left(\frac{\rhovir(z_v)}{4.6 \mearth \, \mathrm{pc}^{-3}}\right)^{1/3} \nonumber\\
&&\times\,
\left[\frac{3.57(-1)^\gamma(\gamma-2)}{B[c(\gamma-2);3-\gamma, \gamma-2]}\right] ,
\eeqa
where $B[z;a,b]$ is the incomplete Beta function.  In Appendix \ref{Sec:concentrations}, we use the findings of the Aquarius simulation \citep{Aquarius08} to derive a relationship between the concentration of local subhalos and their virial mass:
\beq
c = 94 \left(\frac{\mvir}{10^6 \msun}\right)^{-0.067},
\eeq
and we use this relation to determine the subhalo concentration throughout this investigation.

Figure~\ref{Fig:GENimagepath} shows the paths taken by the star's image as the center of a subhalo passes 1 arcsec below the star's position for several values of $\gamma$.  In this figure, $\dl = 50$ pc, $\ds= 5$ kpc, $\vt=200$ km/s, and the subhalo lens has a virial mass of $5\times10^5 \msun$.  We see that the image path depends very strongly on $\gamma$.  If $\gamma\simeq 1$, the motion is nearly linear, and the image moves very little and very slowly.  As $\gamma$ increases, an arc appears in the image path; there is now sufficient mass enclosed in the inner arcsecond to cause an observable vertical deflection when the subhalo passes beneath the star.  Increasing $\gamma$ also increases the motion of the image in a given time period, and the acceleration of the image as the subhalo approaches the point of closest approach becomes apparent.  As the subhalo center passes from the left to the right of star, the star's image will jump from right to left; since the image moves very slowly in the years before and after the crossing of the subhalo, this shift in the image's position offers the best hope for detection.

\section{Observing Strategy}
\label{Sec:strategy}

\begin{figure}
 \centering
 \resizebox{1.0\columnwidth}{!}
  {
       \includegraphics{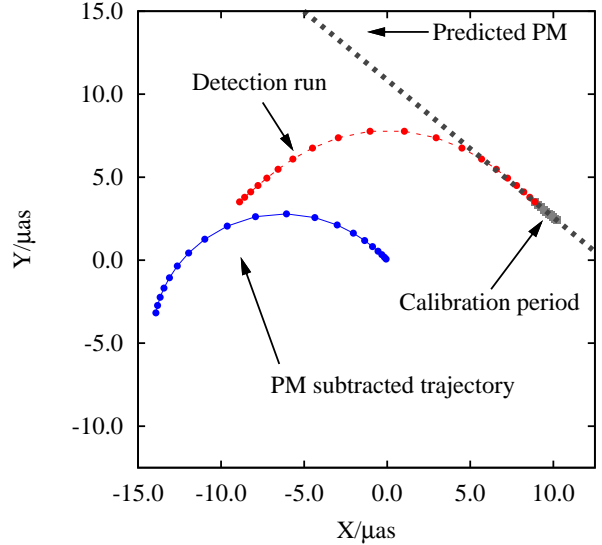}
  }
  \caption{The subhalo lens detection scheme. The dashed line shows the trajectory of an image
induced over 6 years by a subhalo lens with the same properties as 
in Fig.~\ref{Fig:GENimagepath} and $\gamma = 1.8$.   The image motion during the calibration period and the detection run are labeled.  The dotted line shows the direction of the proper motion fitted during the calibration period, and the difference between the measured trajectory and the proper motion prediction is shown by the solid curve.  
For clarity, the
intrinsic proper motion and parallax of the source are not shown.}
 \label{FIG:obs_strategy}
\end{figure}

The image motions shown in Fig.~\ref{Fig:GENimagepath} suggest a simple detection strategy for subhalo lensing, illustrated in Fig.~\ref{FIG:obs_strategy}.   A typical high-precision astrometric search program operates for up to 10 years.  We start with an initial few-year calibration period, during which we assume that the star is relatively far from the lens. During this calibration period we 1) measure the star's intrinsic proper motion, parallax, and starting position \textit{and} 2) search for binary stars or other false positives. Stars that show significant acceleration during the calibration period probably have binary companions, and we reject them from the rest of the search. We follow the calibration period with a several-year detection run, when we essentially wait for a subhalo lens to come close to one of our target stars and induce significant lensing.

The timescale of the calibration period is important, as it needs to
be long enough both to detect binary systems and to obtain a robust
parallax and proper motion measurement.  Increasing the calibration
period length improves the predicted-position accuracy, but it also takes time away from the detection run, reducing the probability of observing a lensing event. The calibration period must be at least one year long to obtain a secure parallax, and we suggest two years as a sensible length to ensure an accurate parallax and proper motion measurement.  The length of the detection run is then set by the duration of the high-precision astrometric campaign. In this analysis, we assume a four-year detection run, implying a total of six years of observations, which is close to the expected mission lengths of Gaia and SIM and a reasonable length for long-term ground-based surveys.  We leave a full discussion of the optimal observing strategy to future work, as the details of the observing scenario will depend on both the particular astrometric technique and the time available for the observations.

For the purposes of this paper, we adopt a simple measurement of the astrometric signal from a lensing event:
\begin{equation}
S = \sqrt{ \sum_{i=1}^{N_\mathrm{epochs}} (Xm_{i} -
Xp_{i})^2 + (Ym_{i} - Yp_{i})^2},
\label{signal}
\end{equation}
where $ Xm_{i}$ and $ Ym_{i}$ give the 2D measured position of the star at each epoch, and $Xp_{i}$ and $ Yp_{i}$ give the 2D position of the star predicted from the proper motion, parallax and position determined during the calibration period. This calculation (following \citealt{Gaudi2005}) essentially measures the total displacement of the star from its expected position over the course of the measurements.  The signal-to-noise ratio (S/N) is simply $S$ divided by the astrometric uncertainty per 1D datapoint ($\sigma$); $\sigma$ includes contributions from the instrument's intrinsic uncertainty per datapoint as well as uncertainty in the star's proper motion, parallax and position. The uncertainty in the star's predicted position grows in time due to the proper-motion uncertainty, and the parallax uncertainty's effect on the 2D measurements varies in direction and magnitude across the sky. 

This calculation gives a conservative estimate of the S/N of a possible lensing detection.  The displacements induced by lensing are all in approximately the same direction, however, and the effective S/N would likely be improved by fitting a model to the data that accounts for these correlations.  We leave such enhancements for future work.

\subsection{False positive removal}

In addition to simply detecting a subhalo lensing signal, we must also
distinguish it from other astrometric signals. Subhalo lensing signals
take place over months, do not repeat, occur without a visible lensing
source, and have a near-unique trajectory. The relatively short event
timescales ensure that our measurements are only sensitive
matter structures on the spatial scales we consider here. The
events' other properties can be used to remove false
positives, such as those generated by stellar microlensing and motion in binary systems.

Astrometric microlensing by a passing point-like lens moves the image centroid along an elliptical path that becomes more circular as the impact parameter increases \citep{Walker95, Paczynski95, Paczynski98, Miralda96,Gould00}.  The image path can become more complicated if the lens has a small finite extent and is opaque \citep{Takahashi03,Lee10}.  In all cases, however, the image completes its orbit on observable timescales, quickly approaching its true position as the compact object moves further away.  We saw in the previous section that subhalos produce a radically different lensing signature; after the passage of the subhalo center, the image moves very slowly and does not approach its true position until the edge of the subhalo passes by the star many decades later.  Lensing by subhalos is therefore readily distinguishable from lensing by compact objects.   

The astrometric signal of a stellar binary is easily distinguished from subhalo lensing signals in most cases simply because the binary system produces a repeating signal.  Almost all long-period systems that do not produce repeating signals in our dataset will be removed by our requirement that the star does not accelerate during the calibration period.  Furthermore, roughly circular binary systems induce a very different astrometric signal from subhalo lenses.

Rare highly eccentric systems with periods much longer than our observation length (or even very close unbound stellar encounters) can produce a short-timescale astrometric signal during periastron, with little signal throughout much of rest of the orbit. The trajectory is similar to a lensing signature (with an additional very large radial velocity signal). However, simple simulations of such systems show that no Keplerian orbit (of any eccentricity $< 0.999$) that produces a lensing-like signal can avoid producing detectable acceleration both during the calibration period and after the putative lensing event.

\subsection{Final subhalo lens confirmation}
The ultimate test of a candidate subhalo is a \textit{prediction} of lensing.   We will show in the next section that detectable dark matter subhalos are probably within 100 pc of our location; while it is possible to detect a more distant subhalo, the subhalo would have to be massive ($\mvir \gsim 10^6 \msun$), and we expect such objects to rarely pass between us and a target star.  Detectable dark matter subhalos are therefore likely to have proper motions greater than half an arcsecond per year, and they will astrometrically affect all stars within several arcseconds of the subhalo center. 
If a halo is detected, its path can be predicted (albeit initially at low precision), and a catalog prepared of stars that are likely to be affected by lensing over the next few years. Intensive astrometric monitoring of faint stars in the field could then provide a fairly rapid confirmation of the existence of the subhalo lens.

\section{Cross Sections for Astrometric Lensing by Subhalos}
\label{sec:x-sections}

\subsection{Signal Calculations}

\begin{figure*}[tb]
 \centering
 \subfigure{
 \resizebox{\columnwidth}{!}
  {
       \includegraphics{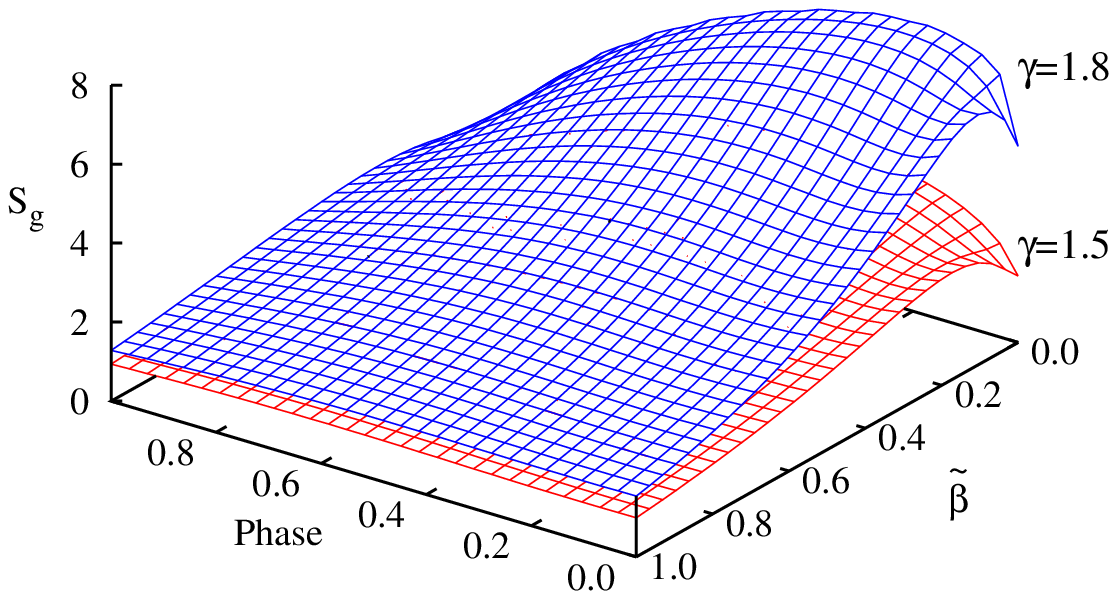}
  }
 }
 \subfigure{
 \resizebox{\columnwidth}{!}
  {
       \includegraphics{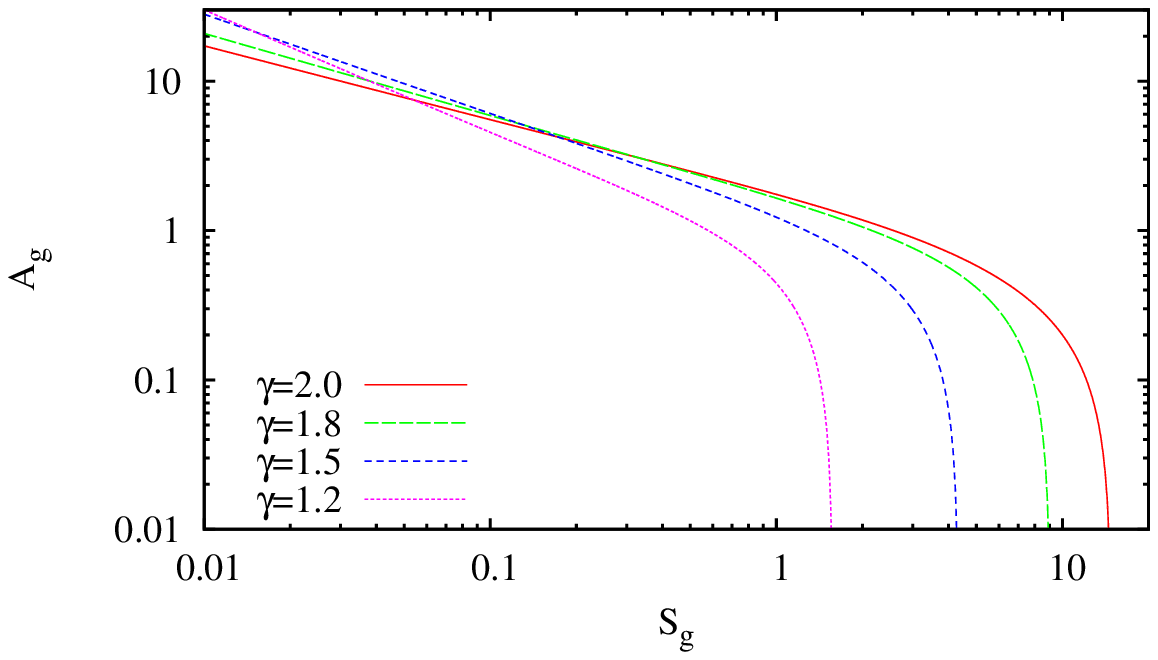}
  }
 }
 \caption{\textit{Left:} The signal $\rm\it S_g \rm$ as a function of phase
($\varphi$) and impact parameter
$\tilde{\beta}\equiv \beta_y/\Delta\beta$ for a
two-year calibration period and a four-year detection run. Note the
decrease in signal toward larger phases, where the image only
partially traverses its lensing path during the observational period.
The decrease in $\rm\it S_g \rm$ at the smallest phases and impact parameters corresponds to lenses that start to produce large image motion during the calibration period, which is then partially subtracted out by the proper motion removal during the detection run.
At larger values of $\tilde{\beta}$ the apparent motions are still
relatively large (see Fig.~\ref{Fig:SISimagegrid}, for instance), but
these motions are difficult
to distinguish from the star's proper motion, leading to the decrease in $\rm\it S_g \rm$.
\textit{Right:} The area $A_g$ in the $\varphi-\tilde{\beta}$ plane that produces a geometrical signal that is larger than a given value for $\rm\it S_g \rm$ for a variety of $\gamma$ values.  Note that the normalized area goes to zero at a value of $\rm\it S_g \rm$ that is dependent on $\gamma$.  For $\rm\it S_g \rm$ values much smaller than this cut-off, $A_g$ is proportional to $({\rm\it S_g \rm})^p$, where the index $p$ depends on $\gamma$.
\label{Fig:sg_3d}
}
\end{figure*}

As described in Section \ref{Sec:strategy}, we calculate the lensing signal by taking the square root of the sum of the squared differences between the star's position on the sky and the position predicted by the proper motion and parallax measured during the calibration period [Eq.~(\ref{signal})].  At each epoch, the difference between the measured image position and the predicted position is proportional to the deflection angle $\vec{\alpha}$.  This linearity implies that we may use any vector $\vec{\eta}$ that is proportional to $\vec{\alpha}$ to calculate the signal; we just have to multiply the resulting signal by $\alpha/\eta$ to obtain the physical signal that would be measured during a lensing event.  We use this technique to separate the signal's dependence on the geometry of the lensing scenario from its dependence on the properties of the lens.  As we describe in detail below, we can then calculate the geometrical signal once and then use that result to determine the signal for any lens.  

Following our convention, we work in a coordinate system where the subhalo's transverse velocity $\vec{v}_\mathrm{T}$ lies along the $x$-axis.  In this case, $\beta_y$ is the fixed impact parameter, and $\beta_{x,0}$ specifies the initial position of the lens.  It is useful to define
\beq
\varphi \equiv \frac{d_L\beta_{x,0}}{\vt t_\mathrm{obs}} \quad \mathrm{and} \quad
\tilde{\beta} \equiv \frac{d_L\beta_{y}}{\vt t_\mathrm{obs}},
\label{geomdefs}
\eeq
where $\beta_{x,0}$ and $\beta_{y}$ are in radians, and $t_\mathrm{obs}$ is the length of the observation (not including the calibration period).  If we define $\Delta\beta$ to be the angular distance, in radians, traversed by the lens during the observational period, we can easily interpret $\varphi$ and $\tilde{\beta}$.  The normalized impact parameter $\tilde{\beta} = \beta_y/\Delta\beta$, while the phase $\varphi =\beta_{x,0}/\Delta\beta$ specifies the location of the point of closest approach to the star along the subhalo's path.  Since we are only interested in cases where the subhalo center passes by the star during the observational period, we constrain $0<\varphi<1$.  
With these definitions Eq.~(\ref{genalpha}) may be rewritten as
\beq
\vec{\alpha} = \theta_\alpha \left(\frac{\vt t_\mathrm{obs}}{r_0}\right)^{2-\gamma}\vec{\eta}\,(\varphi, \tilde{\beta}, t/t_\mathrm{obs},\gamma),
\eeq
which allows us to separate the geometry of the lens-star system (i.e. $\varphi$ and $\tilde{\beta}$) from all of the lens characteristics apart from $\gamma$. From Eq.~(\ref{genalpha}), we see that
\beqa
\vec{\eta} &=&  \hat{\beta}\,\left(\frac{\xi}{\vt t_\mathrm{obs}}\right)^{2-\gamma} \\
&=& \left[{\sqrt{\left({\varphi-\frac{t}{t_\mathrm{obs}}}\right)^2+\tilde{\beta}^2}\,}\right]^{1-\gamma}\left[\left({\varphi-\frac{t}{t_\mathrm{obs}}}\right)\hat{x}+\tilde{\beta}\,\hat{y}\right].
\eeqa

We use $\vec{\eta}$ to calculate the signal of a lensing event.  This is advantageous because the resulting signal is independent of $\theta_\alpha, d_L,$ and $\vt$; we call this signal $\rm\it S_g \rm$ (for ``geometrical signal") because it depends only on $\varphi$, $\tilde{\beta}$, and $\gamma$.  
The calculation of $\rm\it S_g \rm$ takes into account the subtraction of the proper motion and parallax measured during the calibration period, including any apparent motion generated by lensing, as shown in Fig.~\ref{FIG:obs_strategy}.
Figure \ref{Fig:sg_3d} shows $\rm\it S_g \rm(\varphi, \tilde{\beta})$ for $\gamma=1.5$ and $\gamma=1.8$.  We see that $\rm\it S_g \rm$ decreases with increasing impact parameter $\tilde\beta$ and increases with increasing $\gamma$, which is not surprising given the image paths shown in Fig.~\ref{Fig:GENimagepath}.  We also see a preference for geometries in which the subhalo passes by the star earlier in the observational period; the signal is enhanced because there are more epochs after the shift in the star's position when the subhalo center passes the star.

To relate the physical signal $S$ to the geometrical signal $\rm\it S_g \rm$ we use 
\beq
S = \theta_\alpha \left(\frac{\vt t_\mathrm{obs}}{r_0}\right)^{2-\gamma} \rm\it S_g \rm.
\label{Sphys}
\eeq
This relation completes the procedure for determining if a star is detectably lensed by a given subhalo.  Given a specific lens and a minimal detectable value for the signal $\rm\it S_\mathrm{min} \rm$, Eq.~(\ref{Sphys}) may be inverted to obtain the corresponding minimal value for $\rm\it S_g \rm$.  We then determine the area $A_g$ of the region in the $\varphi-\tilde{\beta}$ plane (with $0<\varphi<1$ and $\tilde{\beta}>0$) that produces a geometrical signal that exceeds this minimal value for $\rm\it S_g \rm$.   Finally, we convert $A_g$ to a physical area on the sky that gives $S>\rm\it S_{\mathrm{min}}$, and we call this area the cross-section $A(\rm\it \rm\it S_\mathrm{min} \rm)$; in square radians we have
\beq
A({\rm\it S_\mathrm{min}}) = 2 \left(\frac{\vt t_\mathrm{obs}}{d_L}\right)^2 A_g(\rm \it \rm\it S_g \rm). 
\label{Adef}
\eeq
The factor of 2 accounts for the fact that the stars both above ($\tilde{\beta}>0$) and below ($\tilde{\beta}<0$) the lens are equally deflected, and the two factors of $(\vt t_\mathrm{obs}/d_L)$ follow from the definitions of $\varphi$ and $\tilde\beta$ [see Eq.~(\ref{geomdefs})].  

The geometrical area functions $A_g(\rm\it S_g \rm)$ are shown in the right panel of Fig.~\ref{Fig:sg_3d} for several value of $\gamma$.  As indicated by the $\rm\it S_g \rm(\varphi, \tilde\beta)$ surfaces shown in the left panel of Fig.~\ref{Fig:sg_3d}, the area $A_g$ that produces a geometrical signal larger than a specific $\rm\it S_g \rm$ value depends strongly on $\gamma$ and decreases rapidly with increasing $\rm\it S_g \rm$.   The left panel also shows that $\rm\it S_g \rm$ does not go to infinity as the impact parameter $\tilde\beta$ goes to zero.  Consequently, there is a maximal value of $\rm\it S_g \rm$ that is obtainable for each value of $\gamma$.  At this value of $\rm\it S_g \rm$, the area $A_g$ goes to zero, as shown in the right panel of Fig.~\ref{Fig:sg_3d}.  This maximal obtainable value of $\rm\it S_g \rm$ implies that there is a minimum subhalo mass that is capable of generating a detectable signal, as we will see in the next subsection.   For values of $\rm\it S_g \rm$ that are smaller than half of the maximal possible $\rm\it S_g \rm$ value, $A_g \propto ({\rm\it S_g})^p$, as shown in the right panel of Fig.~\ref{Fig:sg_3d}.  For the four $\gamma$ values we tested, we found that $p=-1/\gamma$.   This simple power-law behavior will be shared by the physical lensing cross sections.

\subsection{Properties of the Lensing Cross Sections}
\label{Sec:crosssectionprops}

The basic shape of the lensing cross sections can be deduced from the left panel of Fig.~\ref{Fig:sg_3d}.  This figure shows that $\rm \it \rm\it S_g \rm$ does not depend strongly on the phase $\varphi$ when $\rm \it \rm\it S_g \rm$ is smaller than the maximum obtainable geometrical signal.  For ${\rm\it S_\mathrm{min}}$ values that correspond to these values of $\rm \it \rm\it S_g \rm$, the lensing cross sections are rectangular.  The width of the rectangle is given by the change in the lens position during the observational period ($\Delta\beta$) because we only consider lensing scenarios with \mbox{$0<\varphi<1$}.   The length of the rectangle is two times the maximum impact parameter that a star may have and still be detectably lensed.  The impact parameters that lie within the lensing cross section are therefore bounded by $|\beta_y| \lsim A({\rm\it S_\mathrm{min}})/(2\Delta\beta)$.  For a lens at a distance of 50 pc with $\vt=200$ km/s, Eq.~(\ref{beta}) tells us that $\Delta\beta \simeq 3$ arcseconds for a four-year observational period, so the maximum impact parameter that produces a lensing signal $S>{\rm\it S_\mathrm{min}}$ is $\beta_y\simeq A({\rm\it S_\mathrm{min}})/(6 \,\mathrm{arcsec}$).  

\begin{figure*}[tb]
 \centering
 \subfigure{
 \resizebox{\textwidth}{!}
 {
      \includegraphics{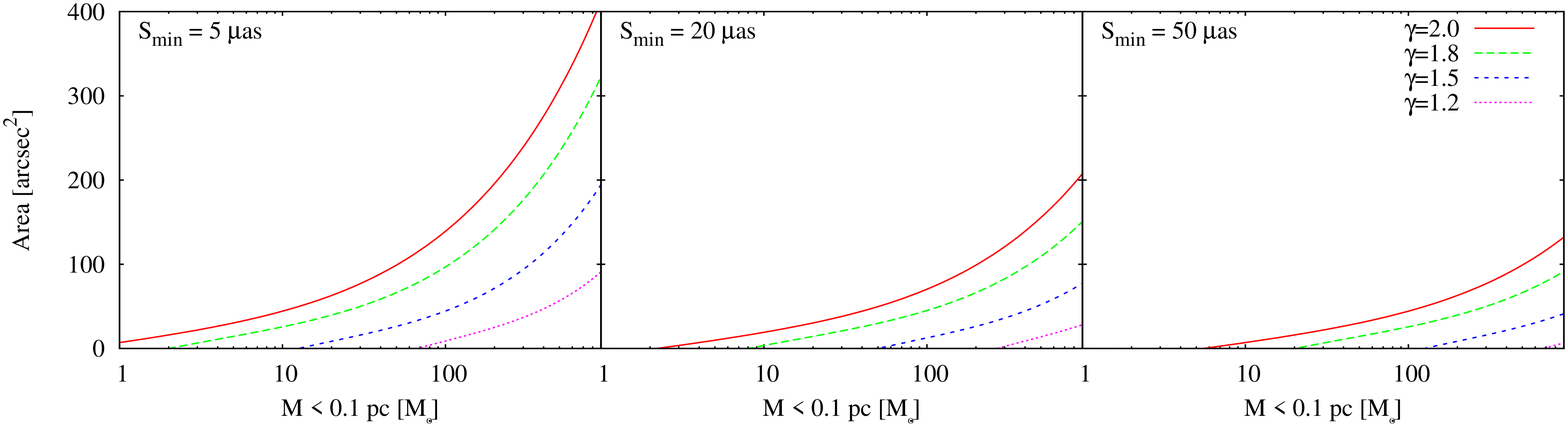}
 }
 }
 \subfigure{
 \resizebox{\textwidth}{!}
 {
      \includegraphics{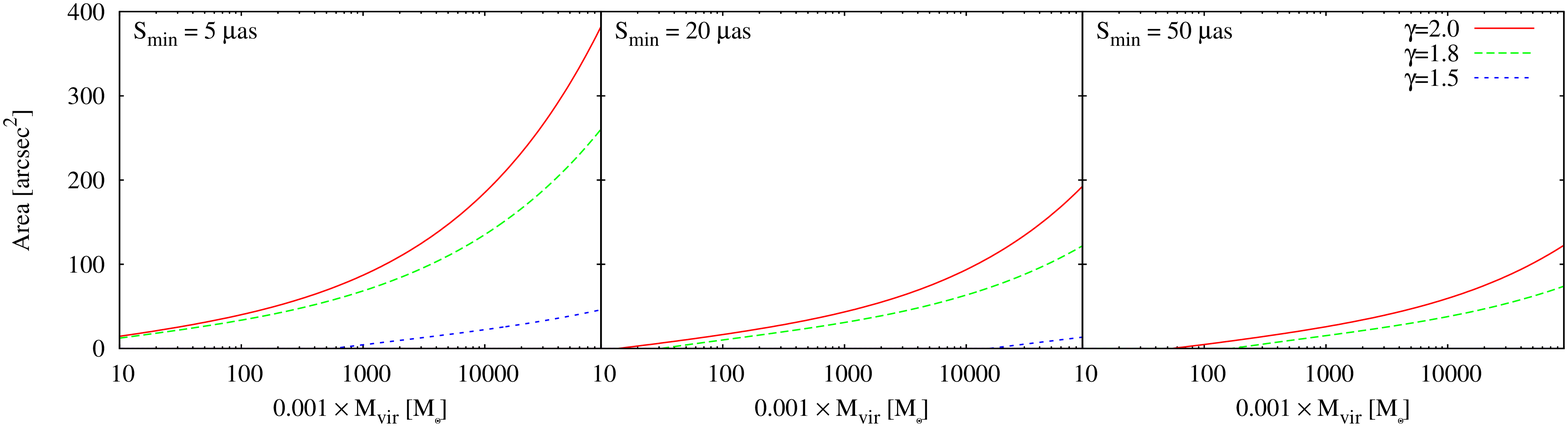}
 }
 }
 \caption{The area around a subhalo that will produce an astrometric signal greater than $\rm\it S_\mathrm{min} \rm$ as a function of subhalo mass, with $\dl=50$ pc, $\ds= 5$ kpc, and $\vt=200$ km/s.
 \textit{Top:} The mass of the subhalo is defined as the mass contained within a radius of 0.1 pc from its center.  Truncating the subhalo density profile at a radius of 0.1 pc does not affect its lensing signature.  \textit{Bottom:} The mass of the subhalo is defined as 0.1\% of its virial mass.  The area curves for $\gamma =1.2$ are not shown because the subhalo must have $\mvir >10^8 \msun$ to generate a signal of at least 5 $\mu$as if $\gamma =1.2$.  \label{Fig:m_area}}
\end{figure*}

Figure \ref{Fig:m_area} shows how the lensing cross section depends on the mass of the subhalo for three values of $\rm\it S_\mathrm{min} \rm$: 5 $\mu$as, 20 $\mu$as, and 50 $\mu$as.  We characterize the mass of the subhalo in two ways, as discussed in Section \ref{Sec:genprofile}. In the top row, we show the cross section as a function of the mass enclosed in the inner 0.1 pc of the subhalo $(\mr)$.  Recall that truncating the subhalo density profile at a radius of 0.1 pc does not significantly alter the lensing signal, which implies that $\mr$ directly determines the lensing signature.  In the bottom row, we show how the lensing cross section depends subhalo's virial mass.  We assume that  99.9\% of the subhalo's virial mass has been lost due to tidal stripping and take $0.001\mvir$ as the present-day mass of the subhalo.   

From Fig.~\ref{Fig:m_area}, we can see how the inner slope of the density profile determines the strength of the lens signature; in all cases, the lensing cross section decreases sharply with decreasing $\gamma$.  The dependence on $\gamma$ is stronger when the virial mass is used to define the mass of subhalo because a shallower density profile requires a larger virial mass to get the same mass within a given radius.  The bottom row of Fig.~\ref{Fig:m_area} indicates that subhalos with $\gamma <1.5$ are not capable of generating easily detectible astrometric lensing signatures, and subhalos with $\gamma = 1.5$ are only detectible if a star passes within a couple of arcseconds of the subhalo center.  The situation is far more promising for subhalos with $\gamma \gsim 1.8$; in this case, an intermediate-mass subhalo could produce a signal of up to 50 $\mu$as if a star passes within 10 arcseconds of the subhalo center.  Moreover, we see that small subhalos ($M_t\lsim 1000 \msun$) are so concentrated that decreasing the inner slope of the density profile from $\gamma=2$ to $\gamma=1.8$ does not significantly change the astrometric lensing signal.  Finally, the top row of Fig.~\ref{Fig:m_area} shows that if the inner regions of the subhalos are denser than predicted by their virial masses, subhalos with shallower density profiles are capable of producing detectible signals. 

For $\rm\it S_\mathrm{min} \rm$ values that correspond to geometrical signals well below the maximal possible value for $\rm\it S_g \rm$ (the value of $\rm\it S_g \rm$ at which $A_g=0$), the cross section has a simple dependence on $\rm\it S_\mathrm{min} \rm$: $A$ is proportional to $({\rm\it S_\mathrm{min}})^p$, with $p=-1/\gamma$ for the four values of $\gamma$ we consider.  This simple power law is directly inherited from the geometrical area functions $A_g(\rm\it S_g \rm)$ discussed in the previous section and shown in the right panel of Fig.~\ref{Fig:sg_3d}.   Consequently, the power index $p$ is independent of $\dl, \ds, \vt$, and subhalo mass, and it is even independent of whether $\mr$ or $\mvir$ is used to characterize the subhalo mass.   As $\rm\it S_\mathrm{min} \rm$ increases toward the maximal possible value, the cross section decreases faster than $({\rm\it S_\mathrm{min}})^p$ and rapidly goes to zero when the maximal possible value for $\rm\it S_\mathrm{min} \rm$ is reached.  

\subsection{Lensing Event Rates}
\label{Sec:event rates}

\begin{figure*}[tb]
 \centering
 \resizebox{\textwidth}{!}
 {
      \includegraphics{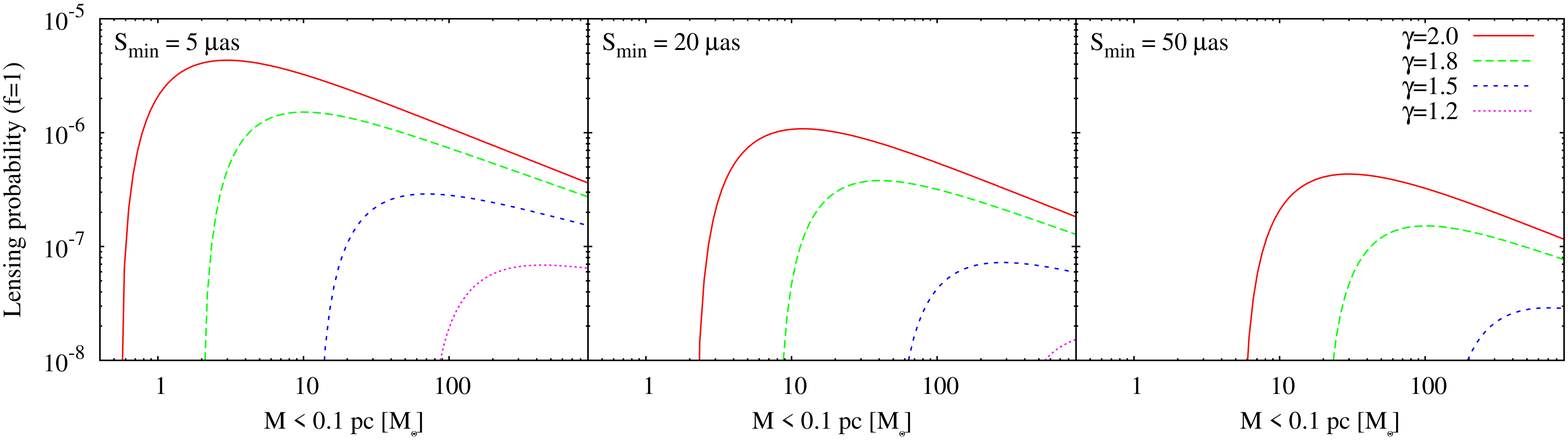}
 }
 \caption{The probability that a star is detectably lensed, with signal $S>\rm\it S_\mathrm{min} \rm$, if all subhalos have the same mass $M$ within the inner 0.1 pc.  The local number density of the subhalos is $f \rho_\mathrm{dm}/M$, where $\rho_\mathrm{dm}$ is the local dark matter density, and $f$ is the fraction of the dark matter contained within 0.1 pc of a subhalo center.  The lensing probability is proportional to $f$, and $f=1$ is shown here to illustrate the maximal possible lensing probability.}
 \label{Fig:probmr}
 \end{figure*}

The lensing cross sections computed in the previous sections may be combined with a subhalo number density to yield a probability that any given star on the sky will be detectably lensed during the observational period.   In this section, we will compute these probabilities using three candidate subhalo number densities.   The first two calculations will assume that all the subhalos have the same mass; we will assume that a fraction $f$ of the local dark matter halo is composed of subhalos with mass $M$ and radii of $R = 0.1$ pc, and then we will assume that the local dark matter halo was once in subhalos with virial mass $\mvir$.   We take the local dark matter density $\rho_\mathrm{dm}$ to be 0.4 GeV cm$^{-3}$, which implies that there is $3.5\times 10^8 \msun$ of dark matter within 2 kpc.  Finally, we will use a subhalo mass function derived from the Aquarius simulations \citep{Aquarius08}.  Throughout this section, we will use $\vt = 200$ km/s and $\tobs =4$ years.

We compute the lensing probabilities by summing the individual cross sections for all subhalos with $\dl<\ds$ for some fixed value of $\ds$.  We choose $\ds$ to be small enough that we may neglect the spatial variation in the subhalo number density within this sphere, and we assume that the subhalos are isotropically distributed.   From Eq.~(\ref{Adef}), we see that the lensing cross section for an individual halo is $A \propto \dl^{-2} A_g$, and $A_g$ depends on $\dl$ only through the $(1-\dl/\ds)$ factor in $\theta_\alpha$.   It follows that the total cross section is 
\beqa
A_\mathrm{tot}&=& 8\pi \,(\vt \tobs)^2 \,\ds \,\nsub(M) \nonumber \\
&&\times\int_0^1 A_g\left[\frac{\rm\it S_\mathrm{min} \rm}{\theta_\alpha(x = \dl/\ds)}\left(\frac{r_0}{\vt \tobs}\right)^{2-\gamma}\right] \drm x,
\label{distint}
\eeqa
where $\nsub(M)$ is the number density of subhalos with mass $M$.   Although $\ds$ will typically vary from less than 1 kpc to \mbox{5 kpc} for stars in a high-precision astrometric survey, we simplify the calculation by assuming that all the monitored stars are 2 kpc from us.   Since $\Atot$ increases linearly with $\ds$, taking $\ds=2$ kpc effectively averages over stars that are uniformly distributed over 1 kpc $<\ds<$ 3 kpc.  

We will see that the resulting total cross section $A_\mathrm{tot}$ is much smaller than the total area of the sky $A_\mathrm{sky}$.  It is therefore highly unlikely that the cross section for lensing by one subhalo will overlap with the cross section associated with a different subhalo, and we may consider $A_\mathrm{tot}$ to be the total area on the sky in which a star would be detectably lensed.   Furthermore, since the individual subhalo lensing cross sections are less than 0.1 square arcminute (see Fig.~\ref{Fig:m_area}), the probability of having multiple stars within the cross section of a particular subhalo is low enough that we may consider each star to be an independent sample of the sky.  In this case, we may interpret the fraction $A_\mathrm{tot}/A_\mathrm{sky}$ as the probability that any single star is detectably lensed by a subhalo.  A subhalo in the Galactic plane is more likely to be detected than a subhalo near the Galactic pole due to the higher density of target stars in the plane, but, since the subhalos are isotropically distributed and can only lens one star apiece, the average lensing probability ($A_\mathrm{tot}/A_\mathrm{sky}$) is individually applicable to each star in the sky.
 
\subsubsection{Event Rates: Mono-mass subhalos}
\label{Sec:event rates mono}

We first consider cases where all the subhalos have the same mass $\mr$ within a radius of 0.1 pc.  The local subhalo number density is then $\nsub = f \rho_\mathrm{dm}/\mr$, where $f$ is the fraction of the dark matter that is contained in the inner 0.1 pc of the subhalos; $f<1$ could mean that there is a smooth component of the local dark matter distribution, or it could mean that the subhalos' radii are larger than 0.1 pc and their actual masses are $\mr/f$.  In Fig.~\ref{Fig:probmr}, we show the lensing probability with $f=1$ for three values of $\rm\it S_\mathrm{min} \rm$: 5 $\mu$as, 20 $\mu$as, and 50 $\mu$as.  We see that the lensing probability is highest for each value of $\gamma$ if the subhalo mass is just slightly larger than the minimum mass required to generate a sufficiently large signal.  Although more massive subhalos have larger cross sections, as shown in Fig.~\ref{Fig:m_area}, the dependence of $A$ on $\mr$ is not steep enough to compensate for the diminishing number density of subhalos as $\mr$ increases.  

If we instead assume that all of the local dark matter was originally in subhalos with virial mass $\mvir$ and that the central regions of these subhalos survive to the present day, then the subhalo number density is $\nsub = \rho_\mathrm{dm}/\mvir$.   As in the previous case, the $(1/\mvir)$ factor in $\nsub$ implies that the lensing probability will peak near smallest value of $\mvir$ that is capable of generating a signal.  For $\rm\it S_\mathrm{min} \rm = 5 \mu$as, the lensing probability peaks at $\mvir \simeq 10^4 \msun$ for $\gamma = 2$ and $\gamma=1.8$, and $\mvir \simeq 10^{6.5} \msun$ for $\gamma=1.5$, as predicted by Fig.~\ref{Fig:m_area}.  Since the cross section for these smaller subhalos does not change much between $\gamma = 1.8$ and $\gamma=2.0$, the lensing probabilities for these two cases are very similar and they peak at probabilities of $5\times 10^{-10}$ and $7\times 10^{-10}$, respectively.  The $\gamma = 1.5$ case peaks a far lower probability of $1\times 10^{-12}$.  It is not surprising that these numbers are about four orders of magnitude lower than the peak probabilities in Fig.~\ref{Fig:probmr}; from Fig.~\ref{Fig:m_area}, we see that $A(\mr) \sim A(\mvir)$ for $\mr\sim 10^{-4} \mvir$ and $\gamma =2$, so by setting $\nsub = \rho_\mathrm{dm}/\mvir$, we are effectively setting $f\sim10^{-4}$ in Fig.~\ref{Fig:probmr}. 

When we assume that all the subhalos have the same mass, then $A_\mathrm{tot}$ has the same simple dependence on $\rm\it S_\mathrm{min} \rm$ as $A$; for $\rm\it S_g \rm$ values that are small compared to the maximum accessible value, $A_\mathrm{tot}$ is proportional to $({\rm\it S_\mathrm{min}})^p$, with $p=-1/\gamma$ as in Section \ref{Sec:crosssectionprops}.  If we instead consider a subhalo mass function and integrate over subhalo masses, the dependence of $A_\mathrm{tot}$ on $\rm\it S_\mathrm{min} \rm$ changes because there is no longer a single value of $\rm\it S_\mathrm{min} \rm$ that corresponds to the value of $\rm\it S_g \rm$ at which $A_g=0$. 

\subsubsection{Event Rates: subhalo mass function}
\label{Sec:event rates aquarius}

To evaluate the lensing probability with a range of subhalo masses, we will use the local subhalo mass function derived in Appendix \ref{Sec:massfunction} from the results of the Aquarius simulations \citep{Aquarius08}:
\beq
\frac{\drm \nsub}{\drm \mvir} = 2 \times 10^{-5} \, \left(\frac{\mvir}{\msun}\right)^{-1.9}.
\label{dndmvir}
\eeq
As described in Appendix \ref{Sec:massfunction}, this subhalo mass function is applicable within a few kpc of the Sun, and we have assumed that the subhalos in this region lose 99\% of their virial mass due to tidal stripping by the smooth component of the dark matter halo and other subhalos (stars are not included).  We evaluate $A_\mathrm{tot}$ by replacing $\nsub$ in Eq.~(\ref{distint}) with ${\drm \nsub}/{\drm \mvir}$ and integrating over $\mvir$ from $M_\mathrm{min}$ to $M_\mathrm{max}$.  For each value of $\gamma$ and $\rm\it S_\mathrm{min} \rm$, there is a minimal value of $\mvir$ needed to make $A_g$ nonzero; this minimal virial mass is always larger than 10$\msun$, so we set $M_\mathrm{min} = 10 \msun$.  We choose $M_\mathrm{max}$ such that the expectation value for the number of 
subhalos with $\mvir\geq M_\mathrm{max}$ within 2 kpc is greater than one; from Eq.~(\ref{dndmvir}), we have $M_\mathrm{max} = 3 \times 10^6 \msun$.   Extending the integral to larger values of $\mvir$ changes the meaning of $A_\mathrm{tot}$; it is no longer a sum of cross sections for the subhalos expected to be within 2 kpc.  Instead, $A_\mathrm{tot}$ would also include the cross sections for subhalos that we do not expect to find within 2 kpc, multiplied by the probability that such a halo is in this volume. 

The value of $M_\mathrm{max}$ determines the shape of the $A_\mathrm{tot}(\rm\it S_\mathrm{min} \rm)$ function because it determines the value of $\rm\it S_\mathrm{min} \rm$ at which $A_\mathrm{tot}$ goes to zero.  If $\rm\it S_g ({\rm\it S_\mathrm{min} }, M_\mathrm{max})$ is greater than the maximum reachable value for $\rm\it S_g \rm$, then no subhalo with $\mvir < M_\mathrm{max}$ can generate a signal and $A_\mathrm{tot} =0$.  For $\gamma =$ 1.2, 1.5, 1.8, and 2.0, this maximum value for $\rm\it S_\mathrm{min} \rm$ is 0.46, 10, 250, and 750 $\mu$as, respectively.  
If $\rm\it S_\mathrm{min} \rm$ is much less than these upper limits, then $A_\mathrm{tot}(\rm\it S_\mathrm{min} \rm)$ is roughly a power law, and the lensing probability is approximately
\beqa
\left.\frac{A_\mathrm{tot}}{A_\mathrm{sky}}\right|_{\gamma=1.8} &=& 8.7\times10^{-12} \left(\frac{\rm\it S_\mathrm{min} \rm}{5 \, \mu\mathrm{as}}\right)^{-1.74}\mathrm{for}\,\rm\it S_\mathrm{min} \rm<80 \,\mu\mathrm{as},\quad \label{Atot18}\\
\left.\frac{A_\mathrm{tot}}{A_\mathrm{sky}}\right|_{\gamma=2.0} &=& 1.3\times10^{-11} \left(\frac{\rm\it S_\mathrm{min} \rm}{5 \, \mu\mathrm{as}}\right)^{-1.44}\mathrm{for}\,\rm\it S_\mathrm{min} \rm<200 \,\mu\mathrm{as}.\quad\label{Atot20}
\eeqa
For larger values of $\rm\it S_\mathrm{min} \rm$, $A_\mathrm{tot}(\rm\it S_\mathrm{min} \rm)$ decreases faster than these expressions and quickly goes to zero at the values listed above.  If $\gamma=1.5$, then $A_\mathrm{tot}/A_\mathrm{sky}=1.4 \times 10^{-12}$ for $\rm\it S_\mathrm{min} \rm=1\,\mu$as, and there is no power-law behavior between $\rm\it S_\mathrm{min} \rm=1\, \mu$as and the zero point at $\rm\it S_\mathrm{min} \rm=10 \,\mu$as.

Including the probabilities that larger subhalos are present within 2 kpc of our location does not significantly affect the total lensing cross section.  If we set $M_\mathrm{max} = 10^{10} \msun$, then the dependence of $A_\mathrm{tot}$ on $S_\mathrm{min}$ is slightly shallower than the power laws given in Eqs.~(\ref{Atot18}) and (\ref{Atot20}), but the differences are not large.  For $S_\mathrm{min} \lsim 20 \, \mu\mbox{as}$, including subhalos with $3 \times 10^6 \msun<\mvir<10^{10} \msun$ increases the lensing probability by less than 20\%.  At the largest $S_\mathrm{min}$ values described by Eqs.~(\ref{Atot18}) and (\ref{Atot20}), extending the integral to these larger subhalo masses increases $A_\mathrm{tot}$ by a factor of three.  The inclusion of larger subhalos has a more pronounced impact on the value of $A_\mathrm{tot}$ for $S_\mathrm{min} \geq$ 0.46, 10, 250, and 750 $\mu$as for $\gamma =$ 1.2, 1.5, 1.8, and 2.0, respectively.  The larger value of $M_\mathrm{max}$ implies that the total cross section does not go to zero at these values of $S_\mathrm{min}$ as it did when we set $M_\mathrm{max} = 3 \times 10^6 \msun$.  This is not an important change, however, because the lensing probability is less than $10^{-13}$ for these large values of $S_\mathrm{min}$.  Moreover, even if it were possible to monitor far more than $10^{13}$ target stars, lensing events would only be observed if a subhalo with $\mvir >3 \times 10^6 \, \msun$ lies between us and the target stars.

\section{Detection Prospects}
\label{sec:detection_prospects}

As discussed in the introduction, it is only recently that astrometric capabilities have advanced to levels where the detection of subhalo lenses becomes possible. In this section we use the results derived above to evaluate current and future astrometric subhalo search techniques. We consider two scenarios: 1) a large-area search for subhalo lenses and 2) a confirmation of a subhalo suspected on the basis of other detection methods.

\subsection{Achievable $\rm\it S_\mathrm{min} \rm$}

We start by evaluating the achievable $\rm\it S_\mathrm{min} \rm$ for an astrometric survey. For a particular astrometric observing strategy, both the number of epochs and the astrometric precision affect $\rm\it S_\mathrm{min} \rm$. For the purposes of this discussion we calculate the best $\rm\it S_\mathrm{min} \rm$ for each technique using the six-year observation setup described in Section \ref{Sec:strategy}. We include the single epoch instrumental uncertainty, as well as a detailed Monte Carlo simulation of the extra per-epoch uncertainties introduced by the measurement and subtraction of the target star's calibration-period position, proper motion and parallax. We marginalize over a wide range of possible parallaxes and proper motions, as well as the full range of possible sky positions.

For a particular observational setup, the simulations produce a scaling factor between the instrumental astrometric uncertainty per epoch ($\sigma_{\rm inst}$) and the final total uncertainty per datapoint ($\sigma$) used to calculate S/N. For the observational setup described in Section \ref{Sec:strategy}, $\sigma = 1.47\sigma_{\rm inst}$. For a typical S/N = 3 detection, an instrument's $\rm\it S_\mathrm{min} \rm$ is then $4.4\sigma_{\rm inst}$. We note that the summed nature of $\rm\it S_\mathrm{min} \rm$ means that astrometric displacements smaller than $\sigma_{\rm inst}$ are indeed detectable in this scheme.

\subsection{Large-Area Surveys}

A practical all-sky search for subhalo lenses requires enough stars that the lensing probabilities described above lead to a significant probability of detection. For example, following the left panel of Fig.~\ref{Fig:probmr}, if all subhalo lenses are singular isothermal spheres and have a $\mr$ of $2\msun$, and our survey is capable of detecting \mbox{$\rm\it S_\mathrm{min} \rm$=5 $\mu$as,} we need to survey $\sim$5$f^{-1}\times10^5 $ stars to have a good chance of detecting a subhalo lens, where $f$ is the fraction of local dark matter that is contained within 0.1 pc of a subhalo center.  Alternately, at an \smin  of 50 \uas we need to survey $\sim$3$f^{-1}\times10^6$ stars.

Ground-based all-sky surveys are currently limited to milli-arcsecond precisions at best (e.g., \citealt{Ivezic2008}), so we do not consider them further here.  The Gaia mission will have an astrometric uncertainty of \mbox{$\sigma_{\rm inst}\simeq35$} microarcsecond per epoch at \mbox{$m_V\simeq12$} for its best targets, with an average of 83 epochs per target\footnote{\url{http://www.rssd.esa.int/index.php?project$=$GAIA\&page$=$Science\_Performance}}.
From a search of the USNO-B1 catalog, we find that $\sim5\times10^6$ stars are covered at this precision \citep{Monet2003}. For these targets, Gaia's \smin is $\sim$260 \uas because coverage of $10^6$ stars requires S/N$\sim$5 to avoid false positives; this precision is too low to detect intermediate-mass subhalos.  Orders of magnitude more stars are covered by Gaia at lower precision, but an extremely large (and very unlikely) subhalo mass would be required to produce detectable lensing at those precisions. Although there remains a small probability that Gaia could see a lens, we conclude that Gaia's astrometric precision is probably insufficient for a useful all-sky subhalo search. That said, it is prudent to at least attempt a subhalo lens search using the Gaia dataset, as the data will be available and can be readily searched for such signals.

For the N-body simulation-based lensing probabilities discussed in section \ref{Sec:event rates aquarius}, where typical lensing probabilities are $\sim 10^{-11}$ at \smin$=5$ \uas, much improved instrumentation capabilities would be required for detection. The only currently planned instrument capable of reaching \smin$=5$ \uas routinely is SIM, but SIM will at most cover tens of thousands of targets during its lifetime \citep{Unwin2008}. If the subhalo mass functions derived from simulations are correct, a SIM-precision all-sky search would have to cover $10^{11}$ targets to have a good chance of making a detection. This capability will most certainly not be available in the near future. However, it is worth stating that a SIM-precision mission covering $10^8$ targets (a possible capability for a next-generation all-sky astrometric survey), would place unique constraints on the subhalo mass function. In particular, such a survey could usefully evaluate if the simulations under-predict the subhalo mass function or if the subhalos are more dense than expected (the scenario in Fig.~\ref{Fig:probmr}).

Similar conclusions can be drawn for blind searches of sub-areas of the sky: current and planned astrometric capabilities are insufficient for a large-area survey for subhalo lenses. However, if a local subhalo is suspected on the basis of other detection methods, targeted surveys are capable of either detecting the lens directly or stringently constraining its properties.

\subsection{Targeted Observations}

If we have some idea of where a lens might be, searching for that lens becomes much easier. For example, it has been suggested that the Fermi Large Area Telescope \citep{Atwood2009} could detect subhalos in gamma-ray emission (e.g., \citealt{Siegal2008, Ando2009, Buckley2010}). Fermi's first point source catalog \citep{Abdo2010} contains a large number of unidentified sources that could be subhalos capable of producing detectable lensing signals  \citep{Buckley2010}.  Sources in the Fermi point-source catalog are localized to 6 arcminutes (median; 95\% confidence) or even 1.5 arcminutes (best 50 sources; 95\% confidence). The lensing search therefore requires coverage of only 0.01 sq. degrees or less. Furthermore, since Fermi has many plausible sources, we can pick the targets with the best likelihood of detection, such as sources close to the Galactic plane with many astrometric target stars. The aim here is to place the best possible limits on the lens properties (with the possibility of an actual detection), and even current techniques (reaching $\rm\it S_\mathrm{min} \rm\simeq50$ \uas) could place useful limits on the lens properties.  If astrometric lensing is detected around a gamma-ray source, then  the magnitude of the deflection provides a measurement of the subhalo's central density; the shape of the image's path provides information about the inner density profile; and the rate of change in the image's position provides information about $\vt/\dl$.  If no lensing is detected, constraints could be placed on a combination of the object's distance, mass, and density profile.

Since it is unlikely that we will know the exact position of a suspected subhalo, moderately wide-field astrometric techniques are favored for this type of search. From space, the Hubble Space Telescope (HST) has demonstrated $\sim$1 mas precision crowded-field astrometry in the cores of globular clusters using the Advanced Camera for Surveys (ACS) instrument (e.g., \citealt{Anderson2010}). However, much better precision has been demonstrated on arcminute scales from the ground. 

AO-equipped 5-10m telescopes routinely achieve 100 $\mu$as precision astrometry \citep{Cameron2009, Lu2009}. These techniques minimize systematics by observing in narrow NIR bands, at consistent air masses, and with careful attention to other sources of systematic error. Such surveys require only a field with several stars within an arcminute field of view (such as is common in the galactic plane) and a few minutes of observing time per field. Using these techniques, relatively small 2m-class telescopes equipped with low-cost adaptive optics systems can reach 50-100 \uas precision in tens of minutes and can perform large, intensive astrometric surveys \citep{Britton2008, Baranec2009, Law2009}. The precision can be further improved; in the absence of systematics, a 10m-class telescope is predicted to reach \mbox{10 \uas} in similar integration times (although currently systematics limit precisions to the $\sim$100 $\mu$as level, development is continuing). On the same basis, a 30m-class telescope could reach few-$\mu$as precision in a few minutes over a small field \citep{Cameron2009}, although the systematic errors are again likely to dominate such observations until technique improvements are made. 

Current ground-based techniques with precisions of \mbox{50-100 $\mu$as} can detect \smin down to $\sim$200 \uas with S/N=3. This precision may be enough to detect nearby large subhalos ($\mvir \gsim 5\times10^7 \msun$ or $\mr \gsim 400 \msun$ for $\gamma\geq1.8$) in the galactic plane with current instruments. If the astrometric accuracy is improved, current 10m-class telescopes could potentially reach \smin$<$50 \uas in 10-minute observations. In crowded regions, such a system performing a long-term astrometric survey could detect subhalos down to stripped masses ($\mbd = 0.001$) of $\sim$1000 $M_{\odot}$ at $\sim$50 pc distances, while 30m-class telescopes could detect subhalos at least an order of magnitude smaller.

SIM offers another possible route for subhalo lens confirmation. The instrument can reach a best precision of \mbox{1 \uas} (and so a detectable \smin of $\sim$5 \uas). Crowding limits require that SIM's target stars are separated by at least $\sim$5 arcseconds from each other, and so the lensing area must subtend at least $\sim$25 sq. arcseconds. With these capabilities, SIM would be capable of confirming a lens 50 pc away down to a stripped virial mass of $\sim$100 $M_{\odot}$ (for $\gamma=$2.0 or 1.8), or equivalently $\mr$ of 10 $M_{\odot}$ (Fig.~\ref{Fig:m_area}). Because SIM is a pointed mission, it can target faint ($m_V$=20) stars at the cost of observing time, making it relatively easy to obtain a sufficient sample of stars near to a putative subhalo position. Although it may not be possible to obtain 1 \uas precision observations of all stars near to a suspected lens because of observation time constraints, we estimate that $\rm\it S_\mathrm{min} \rm$ of 4-18 $\mu$as would be obtainable in modest amounts (weeks) of SIM observation time. SIM observations could thus confirm suspected subhalos near the galactic plane down to stripped subhalo masses ($\mbd = 0.001$) of hundreds of solar masses or $\mr$ of tens of solar masses.

\section{Summary and Conclusions}
\label{sec:summary}
When a dark matter subhalo moves relative to a more distant star, the star's apparent position changes due to gravitational lensing.  By studying the image motion generated by subhalos with isothermal and NFW density profiles, we have determined that the change in the image's position is detectable only if the subhalo's center passes by the star with a small impact parameter ($\lsim 0.01$ pc in the lens plane).  Therefore, only the inner 0.1 pc of a subhalo is relevant for astrometric lensing, and we adopt a general power-law density profile in this region.  We used the findings of the Aquarius simulation \citep{Aquarius08} to derive a relationship between the concentration of local subhalos and their virial masses, which allows us to convert between a subhalo's virial mass and the mass enclosed within 0.1 pc of the subhalo center.  We found that the image paths due to lensing by a subhalo with $\rho\propto r^{-\gamma}$ depend strongly on $\gamma$, with cuspier profiles producing much larger deflections than shallower profiles.  For $\gamma \gsim 1.5$, an intermediate-mass subhalo ($\mvir\gsim10^5 \msun$) within a kpc of the Sun can produce astrometric deflections that are detectable by current and near-future instruments.

We have designed an observing strategy that can be used to detect subhalo lensing in data from typical astrometric surveys (such as Gaia, SIM, or ground-based methods). The setup makes use of the typical subhalo lensing signature: the image starts out essentially
fixed in position, and as the subhalo center passes by, the star rapidly moves to a new position. The star's image then remains nearly stationary for the next several years.  Under our observing scheme, a star's position, proper motion, and parallax is measured during an initial calibration period, and then the star is monitored over the next several years. We define the lensing signal as the difference between the lensed trajectory and the path predicted by the measured proper motion and parallax. This strategy is immune to the most important false-positive possibilities: eccentric binary stars and point-source microlensing.

The magnitude of the resulting astrometric signal depends on the impact parameter between the subhalo and the star.  For a given minimal signal required for detection, a given subhalo will detectably lens all stars within a certain area on the sky.  We computed this cross section for lensing as a function of both the subhalo virial mass and the mass enclosed within \mbox{0.1 pc} for several values of $\gamma$.  Combining these cross sections with a local subhalo number density allows us to calculate the probability that a given star's image will be detectably deflected by a subhalo within a given observation time.  To evaluate the subhalo lensing probability predicted by N-body simulations, we derive a mass function for local subhalos based on the findings of the Aquarius simulation \citep{Aquarius08}.

Finally, we use these cross sections and event rates to evaluate the detectability of subhalo lensing signatures. We consider two scenarios: 1) a large-area survey for subhalo lenses and 2) a confirmation of a subhalo suspected on the basis of other detection methods. We find that Gaia all-sky astrometric measurements are close to being able to constrain subhalos with abnormally high central densities, as could arise if substructure formed very early \citep{RG09, Berezinsky10}.  A subhalo's astrometric lensing signature would also be enhanced if it contains a black hole \citep{BZS05, RG09}; the black hole would steepen the inner density profile and would add a point mass to the subhalo center, resulting in a distinctive astrometric lensing signature.   Given these possibilities, it is certainly prudent to attempt a subhalo lensing search using Gaia, and we leave a thorough investigation of these more exotic scenarios for future work.  Unfortunately, a full-sky survey with much higher astrometric precision than Gaia is required to usefully constrain the dark matter subhalo mass function currently predicted by N-body simulations.

A targeted search for astrometric lensing by subhalos is far more promising;  if a subhalo's presence is suspected by other means (for example, as a Fermi gamma-ray source) current and near-future ground based astrometry surveys are capable of directly searching for the subhalo's lensing signature, down to stripped masses ($0.001\mvir$) of $\sim$1000 $M_{\odot}$ at $\sim$50 pc distances. The SIM astrometric satellite could confirm suspected subhalos near the galactic plane even if the subhalo is 1-2 orders of magnitude less massive or more distant.  Fermi has already observed gamma-ray sources of unknown origin, and the possibility that these sources are dark matter subhalos has been investigated \citep{Buckley2010}.  If Fermi detects a gamma-ray source that is consistent with a nearby intermediate-mass subhalo, then high-precision astrometry could at a minimum constrain the object's distance, mass, and density profile, and it may even provide definitive confirmation for the detection of dark matter substructure.

\acknowledgments 
We thank Niayesh Afshordi, Latham Boyle, Marc Kamionkowski, Annika Peter, and Kris Sigurdson for useful discussions and comments on the manuscript.  

\appendix
\section{Strong lensing and truncation effects}
\label{Sec:multiandtrunc} 
The lensing cross sections presented in Section \ref{Sec:crosssectionprops} confirm that we are firmly in the weak-lensing regime ($\alpha \ll \beta$ in Fig.~\ref{Fig:lensdiagram}).  It follows from Eq.~(\ref{genThetaE}) that the condition $\alpha \ll \beta$ is equivalent to the condition $\theta_E \ll \beta$.  In our coordinate system, with the lens moving along the $x$-axis, the minimal value for $\beta$ is the impact parameter $\beta_y$; we are in the weak-lensing regime only if $\theta_E \ll \beta_y$.   We should excise the area with $\theta_E \geq \beta_y$ from our cross sections because our solution to the lens equation is not valid in this region.   If $\Delta\beta$ is the angular distance traversed by the lens during the observation period, then the area that should be excised is $A_x =2 \Delta\beta \times \theta_E$.  This area is much smaller than the total cross section for lensing,  $A=2 (\Delta\beta)^2 A_g$ from Eq.~(\ref{Adef}),  if $\Delta\beta \gg \theta_E/A_g$.  For all the subhalos that we consider ($\mvir < 10^8 \msun,\, \gamma \leq 2$), $\theta_E/A_g\lsim 10 \,\mu$as.  With this bound and $\vt > 5$ km/s, $\Delta\beta \gg \theta_E/A_g$ is satisfied if $\dl \ll 1000$ kpc.  We therefore conclude that $A_x$ is an insignificant contribution to the lensing cross section and we need not exclude it.

The cross section for lensing also tells us how far a star may be from the center of the subhalo and still be detectably deflected.  As discussed in Section \ref{Sec:genprofile}, we do not truncate the density profile when calculating the surface density of the subhalo, and this approximation is valid only if $\xi$ is much smaller than the truncation radius ($R_t$) of the subhalo.  In Section \ref{Sec:genprofile} we stated that this condition is safely satisfied if $R_t\gsim0.1$ pc, and we will now verify that claim.  We assume that we are only interested in subhalos with virial masses less than $10^8 \msun$ and signals greater than $1\,\mu$as.  These restrictions define a lower bound on detectable values of the geometric signal; for $\gamma = \{2.0, 1.8, 1.5, 1.2\}$, we have $\rm\it S_g \rm > \{0.0020, 0.0056, 0.098, 1.22\}$.  At these small values, $\rm\it S_g \rm$ is nearly independent of the phase $\varphi$, as illustrated in the left panel of Fig.~\ref{Fig:sg_3d}.  Since $\varphi$ is confined to be between 0 and 1, the maximal value of $\tilde\beta$ that keeps $\rm\it S_g \rm$ above these lower bounds is just the area $A_g$ evaluated at the minimal value of $\rm\it S_g \rm$: for $\gamma = \{2.0, 1.8, 1.5, 1.2\}$, we have $\tilde\beta < \{37, 28, 10, 0.2\}$.  From the definition of $\tilde\beta$, we see that $\xi_y = \vt t_\mathrm{obs}\tilde\beta$, and so the maximal distance (in the lens plane) between the star and the subhalo center is $\xi_\mathrm{max} = \vt t_\mathrm{obs}\sqrt{1+\tilde\beta_\mathrm{max}^2}$.  With $ t_\mathrm{obs}= 4$ years, and $\vt=200$ km/s, we have $\xi_\mathrm{max} = \{0.03,0.02, 0.008, 0.0008\}$ pc for $\gamma = \{2.0, 1.8, 1.5, 1.2\}$.  If we take $R_t \gsim 0.1$ pc, then ignoring the truncation of the density profile overestimates $\mtwo(\xi_\mathrm{max})$ by less than 20\% for $\gamma \geq 1.5$ and less than 40\% for $\gamma=1.2$.  Moreover, these are very conservative estimates; for $\mvir \ll 10^8 \msun$ or $\rm\it S_\mathrm{min} \rm \gg 1\, \mu$as, the minimal values of $\rm\it S_g \rm$ will be much larger, leading to smaller values of $\xi_\mathrm{max}$ and less disparity between the truncated and infinite values of $\mtwo(\xi_\mathrm{max})$.
\section{Subhalo Properties}
We are interested in intermediate-mass subhalos $(10 \msun \lsim M_t \lsim 10^6 \msun)$ that are located within a few kpc of the Sun.
Fortunately, numerical simulations have recently reached the resolutions required to probe substructure within a few kpc of the Sun \citep{Aquarius08, ViaLactea208}, with minimum resolvable subhalo masses of $\sim 10^5 \msun$.  In this appendix, we will use the Aquarius simulation results presented by \citet{Aquarius08} to derive a mass function and a concentration-mass relation for these local subhalos.

The Aquarius simulation suite includes simulations of six galaxy-size dark matter halos with $\sim 2\times10^8$ particles in each halo and a higher resolution simulation of one of these six halos (Aq-A) with $1.4\times10^9$ particles.  In the Aq-A simulation, each particle has a mass of 1712 $\msun$, making it possible to identify subhalos with masses greater than $10^5 \msun$.
\citet{Aquarius08} defines the interior of the host halo as a sphere with a mean density that is 50 times the critical density; the radius of this sphere called $r_{50}$, and the mass enclosed is $M_{50}$.  The Aq-A halo has $M_{50}=2.5\times10^{12} \msun$ and $r_{50}=433.5$ kpc. The mass of a subhalo $(M_t)$ is determined by the {\sc SUBFIND} algorithm \citep{Springel01}, which counts the number of gravitationally bound particles and then multiplies by the mass per particle to obtain $M_t$.   We will continue to use $M_t \equiv \mbd \mvir$ to parameterize the effects of tidal stripping on the subhalo's mass. 

\subsection{Subhalo Mass Function}
\label{Sec:massfunction}
The subhalo mass function measured in the Aq-A halo for all subhalos with $r<r_{50}$ is
\beq
\frac{\drm N}{\drm M_t} = a_0 \left(\frac{M_t}{m_0}\right)^{-1.9},
\label{dNdM}
\eeq
with $a_0 = 8.21 \times 10^7 M_{50}^{-1}$ and $m_0 = 10^{-5} M_{50}$.  \citet{Aquarius08} also report that the subhalo number density has the same spatial dependence for all subhalo masses $10^5 \msun \leq M_t \leq10^{10} \msun$:
\beq
n(M,r) = n_0(M) \exp\left[-\frac{2}{\alpha}\left\{\left(\frac{r}{0.46 \,r_{50}}\right)^\alpha -1\right\}\right]
\label{nprop}
\eeq
with $\alpha = 0.678$.  To determine the function $\drm n_0/\drm M_t$, we integrate Eq.~(\ref{nprop}) over $r<r_{50}$ and match the result to Eq.~(\ref{dNdM}).  The resulting mass function is
\beq
\frac{\drm n}{\drm M_t} = \frac{a_0}{1.985\,r_{50}^3} \left(\frac{M_t}{m_0}\right)^{-1.9}\exp\left[-\frac{2}{\alpha}\left\{\left(\frac{r}{0.46 \,r_{50}}\right)^\alpha -1\right\}\right].
\label{genmf}
\eeq

The Aq-A halo is larger than the Milky Way's halo (e.g., \citealt{DMS06, LW08, Xue08, Reid09}), so we must use appropriate values of $r_{50}$ and $M_{50}$ when evaluating Eq.~(\ref{genmf}).  We use the density profile presented by \citet{Xue08} to derive approximate values of $r_{50}$ and $M_{50}$ for the Milky Way: $M_{50} \simeq 9.5 \times 10^{11} \msun$ and $r_{50} \simeq 310$ kpc.  With these parameters,
\beq
\frac{\drm n}{\drm M_t} = \frac{2.5\times10^{-8}}{\mathrm{pc}^3 \msun} \left(\frac{M_t}{\msun}\right)^{-1.9}\exp\left[-\frac{2}{\alpha}\left\{\left(\frac{r}{140 \, \mathrm{kpc}}\right)^\alpha -1\right\}\right].
\label{MWmf}
\eeq
For $r\lsim 20$ kpc, the subhalo number density is no longer strongly dependent  on $r$, and it changes by only 7\% as you move 2 kpc away from the solar radius ($R_0 \simeq 8$ kpc).  We will neglect these small variations so that we may treat the local subhalo number density as isotropic.  If we evaluate Eq.~(\ref{MWmf}) at $r = 8$ kpc, we find
\beq
\frac{\drm n}{\drm M_t} = \frac{3\times10^{-7}}{\mathrm{pc}^3 \msun} \left(\frac{M_t}{\msun}\right)^{-1.9}
\label{dndmt}.
\eeq

We now need to convert this mass function for subhalo mass $M_t$ to a mass function for virial mass $\mvir$.  We will assume that all subhalos within 2 kpc of the Sun lose the same fraction of their mass due to tidal stripping.  If $M_t = \mbd \mvir$, where $\mbd$ is constant, then we have
\beq
\frac{\drm n}{\drm \mvir} = \frac{3\times10^{-7} \, \mbd^{-0.9}}{\mathrm{pc}^3 \msun} \left(\frac{\mvir}{\msun}\right)^{-1.9}.
\label{dndmvirgen}
\eeq
There is great uncertainty surrounding the local value for $\mbd$.  \citet{ViaLactea107} monitored subhalo mass evolution in the Via Lactea simulation, and unsurprisingly found that the fraction of mass lost due to tidal stripping increases closer to the center of the host halo.  They found that subhalos in the region containing the inner sixth of the host halo mass  lose roughly 80\% of their mass between a redshift $z \sim 2$ and the present day.  This sample contains subhalos that are far further from the host's center than the Sun, so we may consider $\mbd \simeq 0.2$ to be a rough upper bound.  Meanwhile, \citet{VTG05} developed a semi-analytical model for tidal stripping and concluded that $0.001\lsim\mbd\lsim0.1$ for all subhalos in a Milky-Way sized host, with most subhalos losing 99\% of their original virial mass.  Finally, Eq.~(\ref{dndmvirgen}) implies that the total mass in a sphere with radius 2 kpc that was once part of a subhalo with $\mvir \lsim 10^8 \msun$  is $6\mbd^{-0.9} \times 10^5 \msun$.  This mass must be less than all the dark matter contained in this sphere ($3.5\times10^8 \msun$), so $\mbd \gsim 0.001$ on average.  We adopt a middle-of-the-road value of $\mbd = 0.01$ for local subhalos when evaluating the lensing event rates in Section \ref{Sec:event rates aquarius}.  This value for $\mbd$ does not include stripping by stellar encounters; here $M_t$ is the subhalo mass measured in N-body simulations that do not include stars.

\subsection{Subhalo Concentrations}
\label{Sec:concentrations}
Many methods for assigning concentrations to dark matter halos have been proposed  (e.g \citealt{Bullock01, Neto07, Duffy08, Maccio08}), 
but these models focus on isolated halos that are far more massive than the subhalos we consider.  Furthermore, numerical simulations indicate that subhalos nearer to the center of the host halo have higher concentrations than both isolated halos \citep{Ghigna98, Bullock01} and subhalos in the outskirts of the host halo \citep{ViaLactea107,Aquarius08, ViaLactea208}.  In light of this distinction, we adopt a $c(\mvir)$ relation for local subhalos that is based on the findings of the Aquarius simulation \citep{Aquarius08}.  Since the virial radius of a subhalo is not easily measured, a different concentration parameter is often used to characterize the concentration of subhalos in simulations:
\beq
\delta_V \equiv \frac{2 V_\mathrm{max}^2}{(H_0r_\mathrm{max})^2},
\label{delV}
\eeq
where $V_\mathrm{max}$ is the maximum circular velocity within the subhalo, and $r_\mathrm{max}$ is the distance from the subhalo center at which the circular velocity is maximized.  The Aquarius team found that $\delta_V$ depends on subhalo mass $M_t$ and distance from the host halo center $r$; when they average over all subhalos with $M_t\gsim3 \times 10^6 \msun$, they find
\beq
\delta_V = 3.8 \times 10^6 \left(\frac{r}{\mathrm{kpc}}\right)^{-0.63},
\label{AqDelR}
\eeq
and when they average over all subhalos, they find\footnote{The exponent is reported incorrectly in the caption of Fig. 28 of \citet{Aquarius08}, but the curve shown in Fig. 28 is correct.}
\beq
\delta_V = 5.8 \times 10^4 \left(\frac{M_t}{10^8 \msun}\right)^{-0.18},
\label{AqDelM}
\eeq
with considerable scatter in both cases \citep{Aquarius08}.  Inspired by these relations, we adopt a model
\beq
\delta_V = N_\delta \left(\frac{r}{\mathrm{kpc}}\right)^{-0.63}\left(\frac{M_t}{10^8 \msun}\right)^{-0.18},
\label{delVmod}
\eeq
and we use the position-dependent subhalo mass function derived in the previous subsection to compare this model with Eqs. (\ref{AqDelR}) and (\ref{AqDelM}).  Matching Eq.~(\ref{AqDelR}) gives $N_\delta = 2.4 \times 10^6$, while matching Eq.~(\ref{AqDelM}) gives $N_\delta = 1.5 \times 10^6$.  Since increasing the subhalo concentration enhances the lensing signal, we adopt the latter value to be conservative.  

Given a full (pre-stripped) density profile for the subhalo, it is possible to relate $\delta_V$ to $c\equiv \rvir/r_{-2}$, where $r_{-2}$ is the radius at which $d\ln \rho/d\ln r = -2$ (e.g., \citealt{ViaLactea107}).  For the density profile given by Eq.~(\ref{GenNFW}),
\beq
\delta_V = \left[\frac{\rhovir(z_v)}{\rho_\mathrm{crit,0}}\right]\left(\frac{c}{y_\mathrm{max}}\right)^3\frac{B[y_\mathrm{max}(\gamma-2); 3-\gamma,\gamma-2]}{B[c(\gamma-2); 3-\gamma,\gamma-2]},
\eeq
where $z_v$ is the redshift at which $\rvir$ is evaluated, $\rho_\mathrm{crit,0}$ is the present-day critical density, $y_\mathrm{max} \equiv r_\mathrm{max} /r_{-2}$, and $B[z;a,b]$ is the incomplete Beta function.  For $1\leq\gamma<2$, $y_\mathrm{max} \simeq 2.1$, and the function $\delta_V(c)$ is not strongly dependent on $\gamma$.  Since we are interested in subhalos with $10^4 \msun<\mvir<10^8\msun$ and $r\simeq 8$ kpc, we only need to consider the range $10^5<\delta_V<10^7$.  In this range, $\delta_V(c)$ is well-approximated by a simple power law:
\beq
\delta_V \simeq 0.049 \left[\frac{\rhovir(z_v)}{\rho_\mathrm{crit,0}}\right]\, c^{2.67}.
\label{cfit}
\eeq

We obtain a final expression for $c(\mvir)$ by inverting Eq.~(\ref{cfit}) and inserting Eq.~(\ref{delVmod}).  With $N_\delta = 1.5 \times 10^6$, we find
\beq
c = 94 \, \mbd^{-0.067} \left(\frac{\rhovir(z_v)}{4.6 \mearth \, \mathrm{pc}^{-3}}\right)^{-0.37} \left(\frac{\mvir}{10^6 \msun}\right)^{-0.067}\left(\frac{r}{8\, \mathrm{kpc}}\right)^{-0.24},
\eeq
where we have defined $\mbd \equiv M_t/\mvir$.  Since $c(\mvir)$ changes  little for $0.01<\mbd<1$, we will take $\mbd =1$ when evaluating the concentration.  The dependence of $c(\mvir)$ on the subhalo's location is also fairly weak; for a fixed value of $\mvir$, $c$ decreases by only 10\% between  8 kpc $<r<$13 kpc and increases by only 25\% between 3 kpc $<r<$8 kpc.  We conservatively take $r=8$ kpc when evaluating the concentration of local subhalos.  Finally, we note that subhalos in simulations that include baryons tend to be more concentrated than subhalos in simulations without baryons \citep{RSHH10}, so it is possible that our model underestimates the concentration of local subhalos.

\bibliographystyle{apj}
\bibliography{arXDMlensing}

\label{lastpage}

\end{document}